\documentclass[longauth]{aa}  

\usepackage{graphicx}
\usepackage{txfonts}
\usepackage{threeparttable}
\usepackage{booktabs}
\usepackage{txfonts}
\usepackage{xcolor}
\usepackage[colorlinks=true,citecolor=blue]{hyperref}
\begin{document} 

   \title{NIKA2 Cosmological Legacy Survey}

   \subtitle{Survey Description and Galaxy Number Counts}

   \author{L.~Bing \inst{\ref{LAM}} 
     \and M.~B\'ethermin \inst{\ref{LAM},\ref{SB}}   
     \and G.~Lagache \inst{\ref{LAM}}
     \and R.~Adam \inst{\ref{Lagrange}}
     \and  P.~Ade \inst{\ref{Cardiff}}
     \and  H.~Ajeddig \inst{\ref{CEA}}
     \and  P.~Andr\'e \inst{\ref{CEA}}
     \and  E.~Artis \inst{\ref{LPSC},\ref{Garching}}
     \and  H.~Aussel \inst{\ref{CEA}}
     \and  A.~Beelen \inst{\ref{LAM}}
     \and  A.~Beno\^it \inst{\ref{Neel}}
     \and  S.~Berta \inst{\ref{IRAMF}}
     \and N.~Billot \inst{\ref{UNIGE}}
     \and  O.~Bourrion \inst{\ref{LPSC}}
     \and  M.~Calvo \inst{\ref{Neel}}
     \and  A.~Catalano \inst{\ref{LPSC}}
     \and  M.~De~Petris \inst{\ref{Roma}}
     \and  F.-X.~D\'esert \inst{\ref{IPAG}}
     \and  S.~Doyle \inst{\ref{Cardiff}}
     \and  E.~F.~C.~Driessen \inst{\ref{IRAMF}}
     \and  D.~Elbaz \inst{\ref{CEA}}
     \and  A.~Gkogkou \inst{\ref{LAM}}
     \and  A.~Gomez \inst{\ref{CAB}}
     \and  J.~Goupy \inst{\ref{Neel}}
     \and  C.~Hanser \inst{\ref{LPSC}}
     \and  F.~K\'eruzor\'e \inst{\ref{Argonne}}
     \and  C.~Kramer \inst{\ref{IRAMF}}
     \and  B.~Ladjelate \inst{\ref{IRAME}}
     \and  D.~Liu \inst{\ref{Garching}}
     \and  S.~Leclercq \inst{\ref{IRAMF}}
     \and  J.-F.~Lestrade \inst{\ref{LERMA}}
     \and  P.~Lustig \inst{\ref{IFPU},\ref{INAFT}}
     \and  J.~F.~Mac\'ias-P\'erez \inst{\ref{LPSC}}
     \and  A.~Maury \inst{\ref{CEA}}
     \and  P.~Mauskopf \inst{\ref{Cardiff},\ref{Arizona}}
     \and  F.~Mayet \inst{\ref{LPSC}}
     \and  A.~Monfardini \inst{\ref{Neel}}
     \and  M.~Mu\~noz-Echeverr\'ia \inst{\ref{LPSC}}
     \and  L.~Perotto \inst{\ref{LPSC}}
     \and  G.~Pisano \inst{\ref{Roma}}
     \and  N.~Ponthieu \inst{\ref{IPAG}}
     \and  V.~Rev\'eret \inst{\ref{CEA}}
     \and  A.~J.~Rigby \inst{\ref{Cardiff}}
     \and  A.~Ritacco \inst{\ref{ENS}, \ref{INAF}}
     \and  C.~Romero \inst{\ref{Pennsylvanie}}
     \and  H.~Roussel \inst{\ref{IAP}}
     \and  F.~Ruppin \inst{\ref{IP2I}}
     \and  K.~Schuster \inst{\ref{IRAMF}}
     \and  A.~Sievers \inst{\ref{IRAME}}
     \and  C.~Tucker \inst{\ref{Cardiff}}
     \and  R.~Zylka \inst{\ref{IRAMF}}}
  \institute{Aix Marseille Univ, CNRS, CNES, LAM (Laboratoire d'Astrophysique de Marseille), Marseille, France
     \label{LAM}
     \and
    Université de Strasbourg, CNRS, Observatoire astronomique de Strasbourg, UMR 7550, 67000 Strasbourg, France
    \label{SB}   
    \and
    Université Côte d’Azur, Observatoire de la Côte d’Azur, CNRS, Laboratoire Lagrange, Nice, France
     \label{Lagrange}
     \and
     School of Physics and Astronomy, Cardiff University, Queen’s Buildings, The Parade, Cardiff, CF24 3AA, UK
     \label{Cardiff}
     \and
     AIM, CEA, CNRS, Universit\'e Paris-Saclay, Universit\'e Paris Diderot, Sorbonne Paris Cit\'e, 91191 Gif-sur-Yvette, France
     \label{CEA}
     \and
     Univ. Grenoble Alpes, CNRS, Grenoble INP, LPSC-IN2P3, 53, avenue des Martyrs, 38000 Grenoble, France
     \label{LPSC}
     \and   
    Max-Planck-Institut fur Extraterrestrische Physik (MPE), Giessenbachstr. 1, D-85748 Garching, Germany
     \label{Garching}
     \and
     Institut N\'eel, CNRS, Universit\'e Grenoble Alpes, France
     \label{Neel}
     \and
     Institut de RadioAstronomie Millim\'etrique (IRAM), Grenoble, France
     \label{IRAMF}
     \and
     Observatoire Astronomique de l’Université de Genève, Chemin Pegasi 51, Versoix, Switzerland
     \label{UNIGE}
     \and
     Dipartimento di Fisica, Sapienza Universit\`a di Roma, Piazzale Aldo Moro 5, I-00185 Roma, Italy
     \label{Roma}
     \and
     Univ. Grenoble Alpes, CNRS, IPAG, 38000 Grenoble, France
     \label{IPAG}
     \and
     Centro de Astrobiolog\'ia (CSIC-INTA), Torrej\'on de Ardoz, 28850 Madrid, Spain
     \label{CAB}
     \and
     High Energy Physics Division, Argonne National Laboratory, 9700 South Cass Avenue, Lemont, IL 60439, USA
     \label{Argonne}
     \and  
     Instituto de Radioastronom\'ia Milim\'etrica (IRAM), Granada, Spain
     \label{IRAME}
     \and
     LERMA, Observatoire de Paris, PSL Research University, CNRS, Sorbonne Universit\'e, UPMC, 75014 Paris, France  
     \label{LERMA}
     \and
     IFPU-Institute for Fundamental Physics of the Universe, Via Beirut 2, 34151 Trieste, Italy
     \label{IFPU}
     \and
     INAF-Osservatorio Astronomico di Trieste, Via Tiepolo 11, 34131 Trieste, Italy
     \label{INAFT}          
     \and
     School of Earth and Space Exploration and Department of Physics, Arizona State University, Tempe, AZ 85287, USA
     \label{Arizona}
     \and
     Laboratoire de Physique de l’\'Ecole Normale Sup\'erieure, ENS, PSL Research University, CNRS, Sorbonne Universit\'e, Universit\'e de Paris, 75005 Paris, France
     \label{ENS}
     \and
     INAF-Osservatorio Astronomico di Cagliari, Via della Scienza 5, 09047 Selargius, IT
     \label{INAF}
     \and    
     Department of Physics and Astronomy, University of Pennsylvania, 209 South 33rd Street, Philadelphia, PA, 19104, USA
     \label{Pennsylvanie}
     \and
    Institut d’Astrophysique de Paris, Sorbonne Université, CNRS (UMR7095), Paris, France
     \label{IAP}
     \and
     University of Lyon, UCB Lyon 1, CNRS/IN2P3, IP2I, 69622 Villeurbanne, France
     \label{IP2I}   
        }

   \date{Received 2023; accepted 2023}

 
  \abstract
   {Finding and characterizing the heavily obscured galaxies with extreme star formation up to very high redshift is key to constrain the formation of the most massive galaxies in the early Universe. It has been shown that these obscured galaxies are major contributors to the stellar mass build-up to z$\sim$4. At higher redshift, and despite recent progress, the contribution of dust-obscured galaxies remains poorly known.}
   {Deep surveys in the millimeter domain are necessary to probe the dust-obscured galaxies at high redshift. We conducted a large observing program at 1.2 and 2\,mm with the NIKA2 camera installed on the IRAM 30-meter telescope. This NIKA2 Cosmological Legacy Survey (N2CLS) covers two emblematic fields: GOODS-N and COSMOS. We introduce the N2CLS survey and present new 1.2 and 2\,mm number count measurements based on the tiered N2CLS observations (from October 2017 to May 2021) covering 1169\,arcmin$^2$. }
   {After a careful data reduction and source extraction, we develop an end-to-end simulation that combines an input sky model with the instrument noise and data reduction pipeline artifacts. This simulation is used to compute the sample purity, flux boosting, pipeline transfer function, completeness, and effective area of the survey (taking into account the non-homogeneous sky coverage). For the input sky model, we used the 117 square degree SIDES simulations, which include the galaxy clustering. Our formalism allows us to correct the source number counts to obtain galaxy number counts, the difference between the two being due to resolution effects caused by the blending of several galaxies inside the large beam of single-dish instruments.}
   {The N2CLS-May2021 survey is already the deepest and largest ever made at 1.2 and 2\,mm. It reaches an average 1-$\sigma$ noise level of 0.17 and 0.048\,mJy on GOODS-N over 159\,arcmin$^2$, and 0.46 and 0.14\,mJy on COSMOS over 1010\,arcmin$^2$, at 1.2 and 2\,mm, respectively. For a purity threshold of 80\%, we detect 120 and 67 sources in GOODS-N and 195 and 76 sources in COSMOS, at 1.2 and 2\,mm, respectively. At 1.2\,mm, the number count measurement probes consistently 1.5 orders of magnitude in flux density, covering the full flux density range from previous single-dish surveys and going a factor of 2 deeper, into the sub-mJy regime. Our measurement connects the bright single-dish to the deep interferometric number counts. At 2\,mm, our measurement matches the depth of the deepest interferometric number counts and extend a factor of 2 above the brightest constraints. After correcting for resolution effects, our results reconcile the single-dish and interferometric number counts, that can be further accurately compared with model predictions.}
   {While having reached its depth for GOODS-N, we expect the final N2LCS survey to be 1.5 times deeper for COSMOS. Thanks to its volume-complete flux selection, the final N2CLS sample will be an ideal reference sample to conduct a full characterization of dust obscured galaxies at high redshift.}


   \maketitle
%

\section{Introduction}

Blind far-infrared (far-IR) to millimeter observations have dramatically improved our understanding of the massive dusty galaxies in the early Universe \citep[e.g.,][]{Blain+02,Lagache+05,Casey+14,Madau+14,Hodge+20}. These sources are believed to be the progenitors of massive quiescent galaxies in dense environments that later emerged at lower redshift \citep{Toft+14,Spilker+19,Valentino+20,Gomez-Guijarro+22b}, and the study on the early phase of their formation and evolution provides crucial tests on the theory of galaxy and structure formation and evolution \citep{Liang+18,Lovell+21,Hayward+21}. Since the start of ground-based (sub)millimeter observations, they became rapidly one of the best ways to find the dusty galaxies at the highest redshift \citep[e.g.][]{Barger+98,Chapman+05,Ivison+07,Hodge+13,Strandet+17,Simpson+20,Dudzeviciute+20}. Contrary to targeted follow-up observations of samples selected at shorter wavelengths \citep[e.g.,][]{Capak+15,Bethermin+20,Bouwens+22}, the dusty galaxy samples from blind far-IR to (sub)millimeter surveys of continuous sky areas are much less affected by a complex selection functions, and thus easier to interpret. There are also statistical studies on DSFGs using the serendipitously detected samples in targeted ALMA observations \citep{Bethermin+20,Gruppioni+20,Venemans+20,Fudamoto+21}. However, these studies are also subject to complex corrections due to clustering and are still limited by the area that could be covered by interferometric observations. \\

Deep and blind surveys are, in particular, ideal to measure the source number counts, which describe the variation of the number density of sources with the source fluxes at given wavelengths. With limited information on individual sources, the number counts still provide constraints on the integrated number density of sources of different fluxes across cosmic time and the selection function is relatively simple to be accounted for in the analysis. Although semi-analytic models with simplified assumptions could make successful predictions on the source number counts, hydrodynamical simulations have been struggling to reproduce this simple observable \citep{Hayward+13,McAlpine+19} and still show tension with observations within certain flux ranges \citep{Lovell+21,Hayward+21}. This indicates that detailed studies on smaller-scale physics, including the spatial distribution of dust and stars, the burstiness of star formation, as well as the initial mass function in (sub)millimeter bright dusty galaxies, are still highly essential to understand the formation and evolution of high-z dusty galaxies \citep{Hodge+20,Popping+20}.\\
    
Due to the limitation of sensitivity or field of view, it is difficult for one blind survey alone to detect a statistically large sample of millimeter sources over a wide range of fluxes and make a complete measurement of the number counts. In practice, the measurement on the number counts of bright millimeter sources, above a few mJy at 1.2\,mm, are predominantly contributed by single-dish observations \citep[e.g.][]{Lindner+11,Scott+12}. On the contrary, ALMA brings most of the constraints on the faint-end number counts at the sub-mJy regime, where single-dish surveys start to be limited by their sensitivity and source confusion \citep[e.g.][]{Fujimoto+16,Gonzalez-Lopez+20,Gomez-Guijarro+22}. Most of the previous studies directly combine the two different types of observations in the interpretation and model comparison. However, it has also been shown that single-dish and interferometer surveys do not provide completely equivalent flux measurements \citep{Hayward+13,Cowley+15,Scudder+16,Bethermin+17}. The higher resolution of interferometers gives a flux estimate for individual galaxies, while the relatively large beam of single-dish observations could introduce additional contribution from close-by faint galaxies to the measured fluxes on the brightest "isolated" galaxy in the beam \citep{Bethermin+17}. Previous studies lack realistic estimate of this effect based on real data from blind surveys. Its impact on the joint analysis of raw single-dish and interferometer number counts is seldom considered.\\

The New IRAM KIDs Array, NIKA2 \citep{Monfardini+14, Calvo+16, Bourrion+16} offers a new promising path for statistical studies of the early stage of galaxy evolution obscured by dust. NIKA2 is a continuum instrument installed on the IRAM 30-meter telescope in October 2015 \citep{Adam+18, Perotto+20}. It allows observations within a 6.5' diameter instantaneous field of view using three detector arrays in two bands simultaneously. These include two arrays with 1140 detectors at 1.2\,mm (255\,GHz), as well as another array with 616 detectors operating at 2\,mm (150\,GHz). Thanks to the large collecting area and the large number of detectors filling a large instantaneous field of view, the combination of the 30m telescope and NIKA2 offers capabilities of sensitive and efficient blind surveys of high-z dusty star-forming galaxies (DSFGs) with an angular resolution of 11.1" and 17.6" at 1.2 and 2\,mm, respectively \citep{Perotto+20}. This is the purpose of the NIKA2 Cosmological Survey (N2CLS). With 300\,h of guaranteed-time observations, N2CLS performs deep blind mappings in the GOODS-N and COSMOS field to make a systematical census on DSFGs from cosmic noon up to the first billion years of the Universe with both large area coverage and unprecedented depth among single-dish millimeter observations. The observations at these relatively long wavelengths are also expected to favor the selection of DSFGs at higher redshift \citep{Blain+02,Lagache+04,Casey+14,Zavala+14,Bethermin+15b}.\\

 In this paper, we introduce the N2CLS survey and present the new 1.2 and 2\,mm number counts measurements based on the tiered N2CLS observations over 1169\,arcmin$^2$ in GOODS-N and COSMOS. Data prior to May 2021 are used. At this date, GOODS-N data were completed, while COSMOS was still being acquired.
 Our measurements already cover an unprecedentedly wide range of source fluxes from one single-dish instrument and consider the impact of the beam in number counts measurements for the first time. This is achieved using a dedicated end-to-end simulation based on the SIDES-UCHUU model \citep{Gkogkou+22}. The paper is organized as follows. In Sect.~\ref{sc:survey_data_red_source_ext} we introduce the survey strategy, present the N2CLS maps and the method of source extraction and photometry. In Sect.~\ref{sc:flux_bias_corr} we describe the framework for the correction of the bias in source detection and flux measurement due to instrument noise, pipeline transfer function, and large beam. Sect.~\ref{sc:numbercounts} presents the 1.2 and 2\,mm source number counts measurements and its comparison with previous observations. We also determine the galaxy number counts, that are derived from the source number counts and our end-to-end simulation based on realistic sky simulations, including the clustering. These galaxy counts are finally compared with model expectations. 
In Sect.~\ref{sc:discussion} we discuss the modelling of the millimeter number counts and the impact of spatial resolution on flux measurements
in single-dish observations. We finally conclude in Sect.~\ref{sc:conclusion}. 

\section{Survey description, data reduction and source extraction}\label{sc:survey_data_red_source_ext}


\subsection{Survey design and observation}\label{sc:survey_design_obs}

The N2CLS was designed to have good statistics on both faint and bright sources through a narrow and deep observation in the GOODS-N field and a wider and shallower observation in the COSMOS field. In GOODS-N, the survey time was chosen to approach the source confusion limit of the IRAM 30m telescope at 1.2\,mm on an area of $\sim$160 arcmin$^2$. Source confusion is the contribution to noise in an image due to the superimposed signals from faint unresolved source at the scale of the observing beam. It was estimated using the model from \citet{Bethermin+12} and considering a source density of 1/20$\Omega$. Values (5$\sigma_{\mathrm{conf}}$) are about 0.68\,mJy and 0.23\,mJy at 1.2 and 2\,mm, respectively, for FHWM of 12 and 18''.
In COSMOS, N2CLS covers a much larger area of $\sim$1000 arcmin$^2$ with a shallower depth, to get a larger sample of brighter sources, biased towards higher redshifts \citep[which is counter intuitive, see][]{Bethermin+15b}. Thanks to the dual band coverage of NIKA2, we simultaneously obtain 1.2 and 2\,mm data from the N2CLS observations.\\

N2CLS observations started in October 2017 and finished in January 2023, under project ID 192-16. For the work presented here, we use 170.85\,h on-field observations in total, that were conducted from October 2017 to March 2021.
 They represent 86.15\,h on GOODS-N and 84.7\,h on COSMOS. For the GOODS-N observations, we executed two groups of 12.0'$\times$6.3' and 6.5'$\times$12.3' scans in orthogonal directions centered on RA=12:36:55.03 and Dec=62:14:37.59. For the COSMOS field, we carried out two groups of 27.0'$\times$34.7' and 35.0'$\times$28.0' orthogonal on-the-fly scans centered on RA=10:00:28.81 Dec=02:17:30.44. The two groups of orthogonal scans in both fields were taken with equal times. In GOODS-N, we made the two groups of scans with a speed of 40 and 35\, arcsec/sec, and position angles of -40$^o$ and -130$^o$ in the RA-DEC coordinate system of the telescope. For COSMOS, the two groups of scans were observed with a speed of 60\,arcsec/sec at position angles of 0$^o$ and +90$^o$ in the RA-DEC coordinate system of the telescope.\\
Observations were conducted with a mean line-of-sight opacity $\frac{\tau_{225\,GHz}}{sin(el)}$ of 0.27 and 0.25 for GOODS-N and COSMOS, respectively, {where $\tau_{225\,GHz}$ is the zenith opacity deduced from the Pico Veleta tau-meter measurement at 225\,GHz.} 

\subsection{Data reduction}\label{sc:data_red}

The N2CLS data are reduced using the "may21a"\footnote{\url{https://www.iram.fr/~gildas/dist/archive/piic/piic-exe-may21a.tar.xz}} version of the PIIC data reduction pipeline developed and supported by IRAM \citep{Zylka13}. Our data reduction script is adapted from the default template provided with the PIIC software. We use the options "deepfield" and "weakSou", which are designed to recover faint sources from the NIKA2 timeline data without prior information on source positions. We used the PIIC iterative procedure with 5 iterations to recover the bright source fluxes by subtracting an estimate of the sky (so-called "source map") constructed by thresholding the previous iteration map at 4$\sigma$. The procedure converges rapidly and 5 iterations are sufficient. At each iteration, following the default PIIC parameters, the emission of the sky is subtracted from the timeline using the 16 best correlated KIDs. The signal is also corrected from sky opacity using the IRAM's Pico Veleta tau-meter measurements extrapolated to the NIKA2 bands.
Additionally, for GOODS-N and COSMOS, we set the order of polynomial function used for baseline correction (parameter "blOrderOrig" in PIIC) to 10 and 17, respectively, which remove residual left-over fluctuations in the signal map (as atmospheric and electronic residual fluctuations). All of the data from array\,1 and array\,3 are reduced together to produce a single 1.2\,mm map, while the data from array\,2 are used to produce the 2\,mm map. With its optimized re-gridding strategy applied to redistribute the KIDs signal, 
PIIC samples properly the Gaussian response of point-like sources with 3'' and 4'' sized pixels at 1.2 and 2\,mm.
The resulting FWHM of point source in the final 1.2 and 2\,mm maps are 11.6" and 18.0", respectively. 

A scan selection is performed by PIIC before the data reduction, which automatically discard the bad scans based on their noise properties (e.g. higher noise linked to weather instabilities). 
We have in total 762 and 394 scans for GOODS-N and COSMOS, respectively. Each scan generates three data files with one for each array. At 1.2\,mm (2 arrays), among 1524 and 788 data files on GOODS-N and COSMOS, PIIC finally keeps 1281 and 700 files for map making. It represents a total on-source time of 78.0\,h and 78.7\,h in GOODS-N and COSMOS, respectively. At 2.0\,mm, 599 and 351 data files from GOODS-N and COSMOS observations are used to produce the maps, respectively. They correspond to 72.8\,h and 79.0\,h on-source time in the two fields. The 225\,GHz zenith opacities are ranging from 0.06 to 0.91 and elevations from 20$^\circ$ to 73$^\circ$. Median opacities are equal to 0.2 for both fields. \\

\begin{figure*}[tb]
   \centering
      \includegraphics[clip=true,width=.99\textwidth]{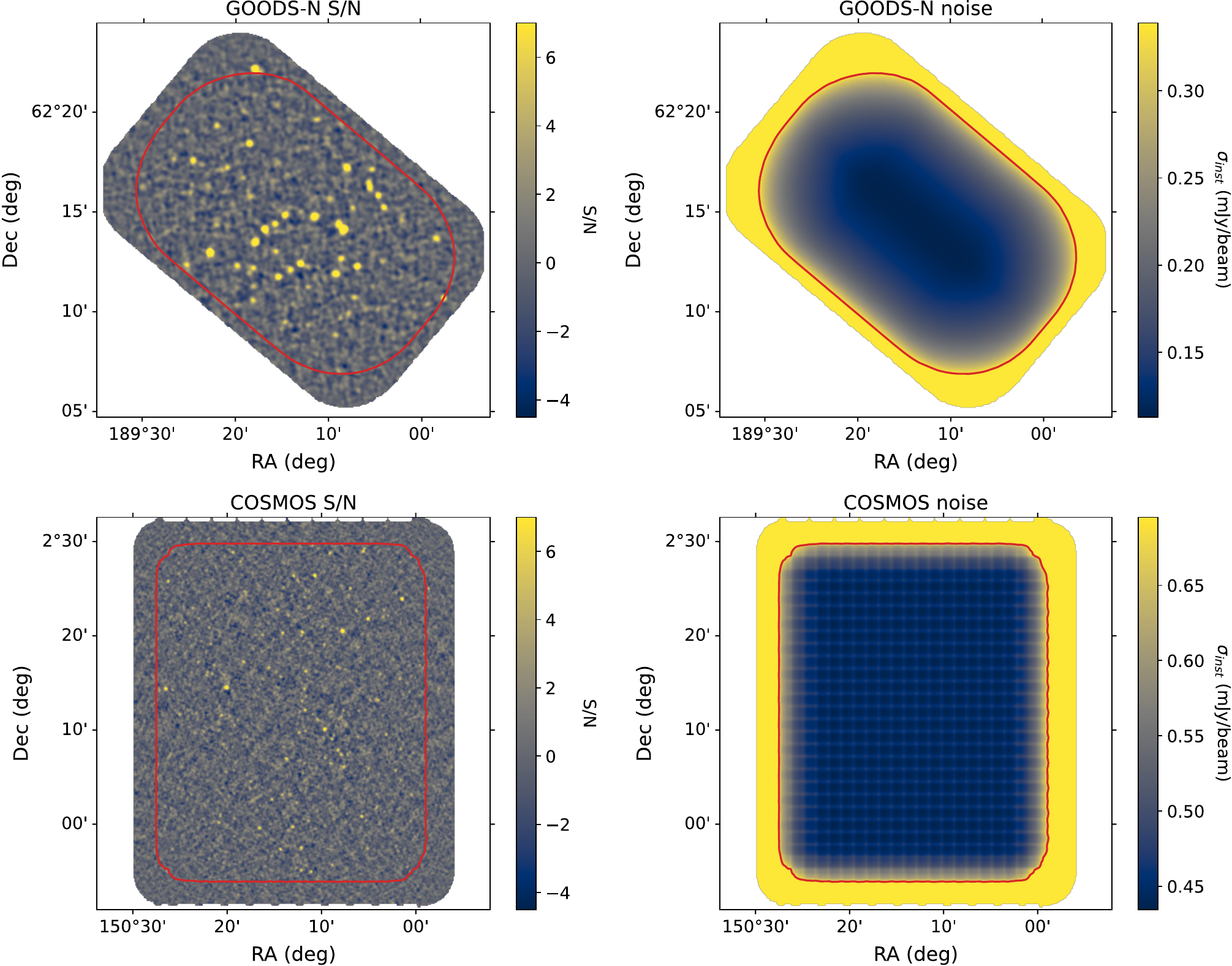}
      \caption{Maps of signal-to-noise ratio (left panels) and noise ($\sigma_{inst}$ in mJy/beam, right panels) of the 1.2\,mm N2CLS maps of the GOODS-N (upper panels) and COSMOS (lower panels) fields. The S/N maps and noise maps are matched filtered (see Sect.~\ref{sc:source_ext}). The regions enclosed in the red contours (159\,arcmin$^2$ for GOODS-N and 1010\,arcmin$^2$ for COSMOS) have sufficiently low noise to be analyzed by our source detection algorithm, and our catalogs and number counts are derived only in these areas (see Sect.~\ref{sc:source_ext} and Sect.~\ref{sc:numbercounts}).}
      \label{fig:n2cls_1.2mm_maps}
   \end{figure*}

\begin{figure*}[tb]
   \centering
      \includegraphics[clip=true,width=.99\textwidth]{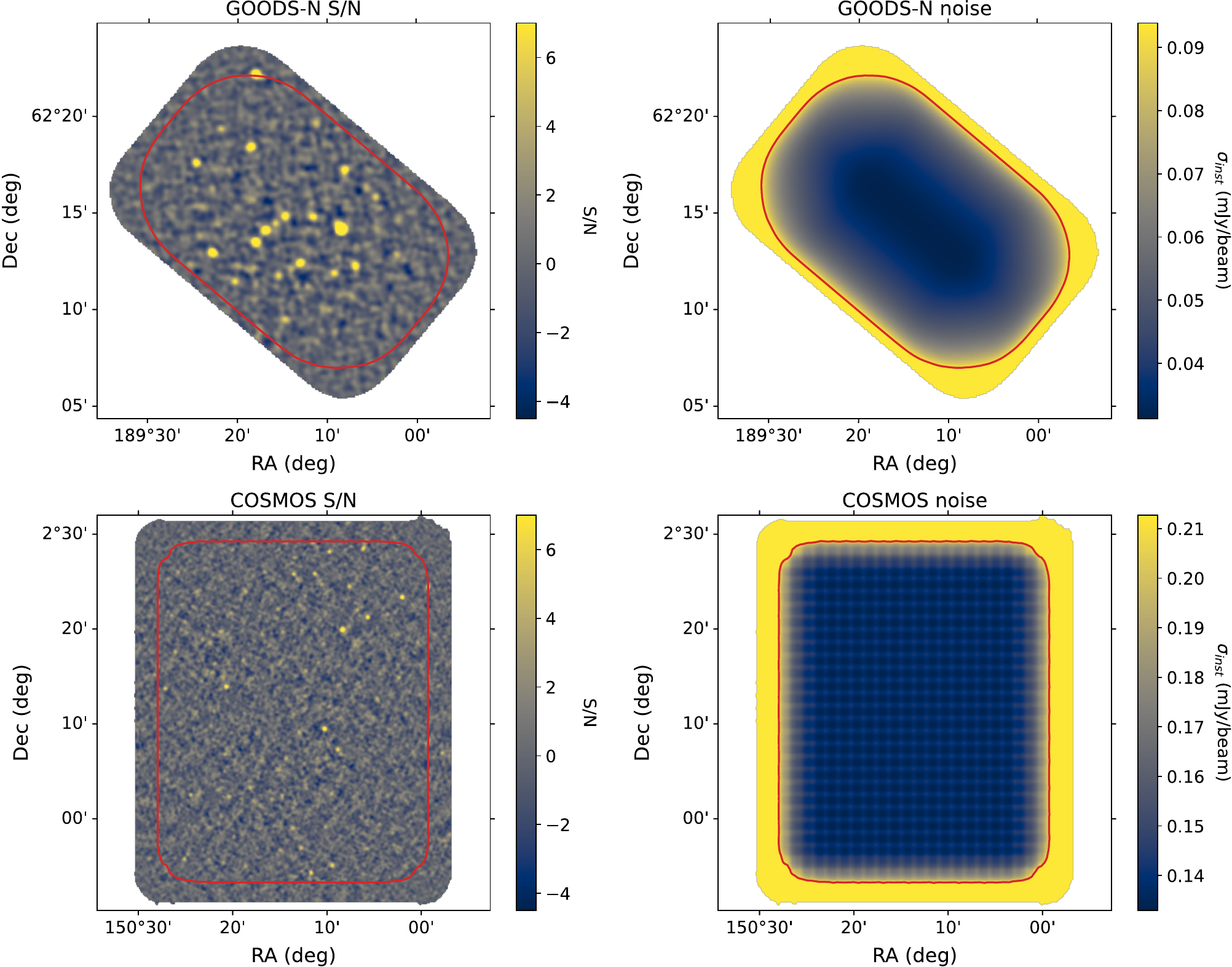}
      \caption{Same as Fig.~\ref{fig:n2cls_1.2mm_maps} but at 2\,mm.}
      \label{fig:n2cls_2.0mm_maps}
   \end{figure*}

In Fig.~\ref{fig:n2cls_1.2mm_maps} and Fig.~\ref{fig:n2cls_2.0mm_maps}, we show the 1.2 and 2\,mm signal-to-noise ratio (S/N) maps of the N2CLS survey in GOODS-N and COSMOS that have been match-filtered by the beam in the corresponding band. The instrument noise maps are also generated from the weight maps produced along with the signal maps, and are also presented in Fig.~\ref{fig:n2cls_1.2mm_maps} and Fig.~\ref{fig:n2cls_2.0mm_maps} for 1.2 and 2\,mm observations, respectively. Considering the high noise level on the edges, our study is restricted to the high-S/N regions delimited by the red lines. These regions are defined as having an instrument noise ($\sigma_{inst}$) smaller than 3 and 1.6 times the minimal value at the center of the GOODS-N and COSMOS field, respectively. This choice is led by a compromise between homogeneity and maximizing the survey area.
The high-quality regions used in our analysis cover 159\,arcmin$^2$ and 1010\,arcmin$^2$ in GOODS-N and COSMOS, respectively. \\

\begin{figure*}[tb]
   \centering
      \includegraphics[clip=true,width=.99\textwidth]{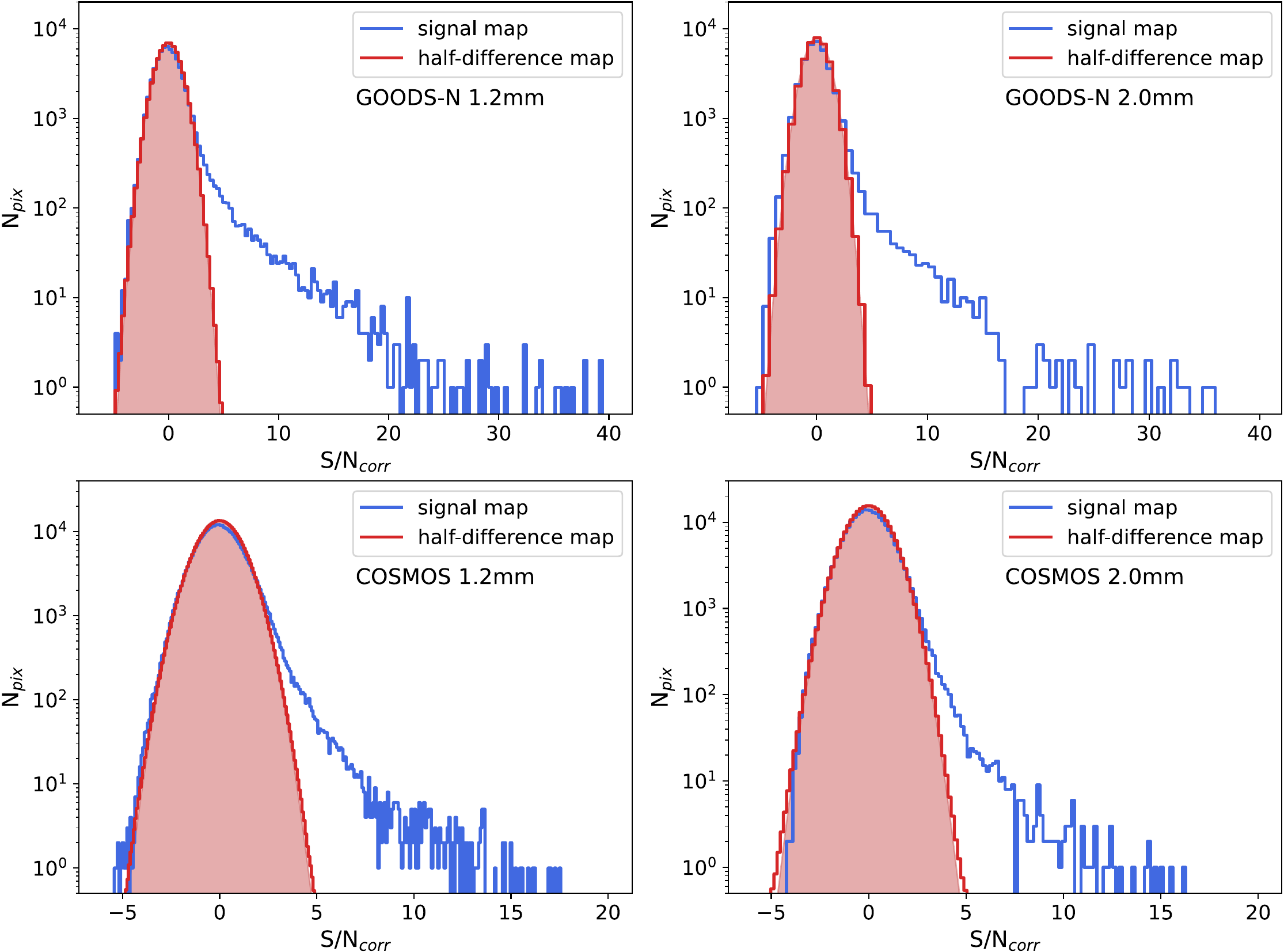}
      \caption{Pixel S/N distribution within the high-quality regions of the 1.2 and 2\,mm maps shown in Figs.~\ref{fig:n2cls_1.2mm_maps} and ~\ref{fig:n2cls_2.0mm_maps}, as well as the average distribution of pixel S/N in 100 randomly generated half-difference maps for each field and waveband (see Sect.~\ref{sc:data_red}). The red shaded region illustrates the best fit normal distribution on the average histograms of the half-difference maps.}
      \label{fig:snr_distribution}
   \end{figure*}

In Fig.~\ref{fig:snr_distribution}, we present the distribution of the pixel values of the S/N map within the high-quality region of each field and each band. The S/N histograms reveal positive tails of high S/N pixels, which indicates that sources are detected by the N2CLS survey. In Table~\ref{tab:tab-surveys}, we provide the central instrument noise level of maps for each field and band. As the noise in the map is not uniform (especially for GOODS-N), we also provide the average instrument noise level in Table~\ref{tab:tab-surveys}. In the high-quality regions of GOODS-N, the 1.2 and 2\,mm maps have average noise levels of 0.17\,mJy and 0.048\,mJy, respectively. For the COSMOS field, we have average noise levels of 0.46\,mJy and 0.14\,mJy within in the high-quality regions at 1.2 and 2\,mm, respectively.\\ 

In Table\,\ref{tab:tab-surveys}, we also compare the noise levels of N2CLS with those of previous surveys. To compare the RMS with surveys at different wavelength than the N2CLS, we rescaled it assuming a far-IR SED template of a typical star-forming galaxy at z=2 from \citet{Bethermin+15}. In GOODS-N, N2CLS is surpassing the depth of any other single-dish millimeter surveys with a similar size for wavelength longer than 1\,mm \citep{Perera+08,Lindner+11,Staguhn+14}. It currently matches the deepest SCUBA2 850\,$\mu$m survey \citep{Cowie+17} after taking into account the SED correction. As for COSMOS, N2CLS is 2.7 and 2.2 times deeper than AzTEC \citep{Aretxaga+11} and MAMBO \citep{Bertoldi+07} at 1.1/1.2\,mm, respectively, and 1.6 times deeper than GISMO at 2\,mm on a 4 times larger area \citep{Magnelli+19}. Similarly to GOODS-N, it also reaches a depth comparable to the deepest SCUBA2 observation at 850\,$\mu$m (S2COSMOS, see \citealt{Simpson+19}). \\

To obtain realizations of the instrument noise as close as possible as in the observations, we generate half-difference maps from the original scans. The half-difference maps, or null maps, are built by opposing
half of the single-scan maps, i.e. multiplying them by $-1$, and co-adding all of them. The opposed single scan maps are selected randomly from the full list. The random selection and coadding operations are carried out by the HalfDifference module in \textit{nikamap} \citep{beelen_alexandre_2023_7520530}\footnote{Available at https://gitlab.lam.fr/N2CLS/nikamap}. This process removes the astrophysical signal and preserves the instrument noise properties if the noise has symmetric properties. At first order this hypothesis is valid with respect to atmospheric fluctuations, tuning variations, and electronic noise or even magnetic fields induced by the telescope, but could be challenged by beam distortions from telescope geometry or differential acceleration during the scans. We will neglect the later as we could not characterize their potential effects on the half-difference maps. Note that there is also a small potential bias due to the weight of each individual maps, which could slightly favor positive or negative signal. But giving the large number of individual scans and generating several realizations, this weight imbalance is minimized.
The distribution of S/N in half-difference maps is shown in Fig.~\ref{fig:snr_distribution} along with its best fit by a normal distribution.

\begin{table*}
\centering
\caption{Comparison of N2CLS-May21 depth to other single-dish (sub)mm surveys in GOODS-N and COSMOS, which are all given as 1$\sigma$ of the noise level. The average and central noise of N2CLS are provided as values in and out parentheses, respectively. For surveys not exactly observing at 1.2\,mm (i.e. AzTEC observations at 1.1\,mm), the root mean square (rms) noise is normalized to 1.2\,mm assuming the average IR SED of star-forming galaxies at z=2 from \cite{Bethermin+15}. Under this assumption, the depth of AzTEC and SCUBA2 observations are divided by a factor of 1.25 and 2.10, respectively. $\nu_{obs}$ is the central frequency of the instruments.}
\label{tab:tab-surveys}   
\begin{threeparttable}
\begin{tabular}{llccc}
\hline
Field & Survey & $\nu_{obs}$ & Area & rms \\
 & & GHz & arcmin$^2$ & mJy/beam \\
\hline 
GOODS-N 0.85/1.2\,mm & AzTEC/JCMT\tnote{1} & 273 & 245 & 0.84 \\
& SCUBA2/JCMT\tnote{2} & 353 & $\sim$140 & $\sim$0.16 \\
& N2CLS-May2021 & 255 & 159 & 0.11(0.17) \\
\hline
GOODS-N 2\,mm & GISMO/30m\tnote{3} & 150 & 31 & 0.14 \\
& N2CLS-May2021 & 150 & 159 & 0.031(0.048)  \\
\hline
COSMOS 0.85/1.2\,mm & AzTEC/ASTE\tnote{4} & 273 & 2592 & 1.00 \\
& MAMBO/30m\tnote{5} & 255 & 400 & 1.00 \\
& S2COSMOS/JCMT\tnote{6} & 353 & 5760 & $\sim$0.47  \\
& N2CLS-May2021 & 255 & 1010 & 0.43(0.46) \\
\hline
COSMOS 2\,mm& GISMO/30m\tnote{7} & 150 & 250 & 0.23 \\
& N2CLS-May2021 & 150 & 1010 & 0.13(0.14) \\
\hline
\end{tabular}
   \begin{tablenotes}
    \item[1] \citet{Perera+08}
    \item[2] \citet{Cowie+17} (Central region)
    \item[3] \citet{Staguhn+14}
    \item[4] \citet{Aretxaga+11}
    \item[5] \citet{Bertoldi+07}
    \item[6] \citet{Simpson+19}
    \item[7] \citet{Magnelli+19}
   \end{tablenotes}
\end{threeparttable}
\end{table*}

\subsection{Source extraction}\label{sc:source_ext}

The source detection of N2CLS survey is made on the matched-filter PIIC maps produced by the dedicated python package \textit{nikamap}. For the matched filter, the kernels are fixed to circular 2D Gaussian with a FWHM equal to the corresponding beam sizes. The matched-filter map absolute level is undefined, and any baseline residual in the PIIC maps could introduce a shift in the matched-filter signal map. Moreover the PIIC re-gridding introduces a correlation in the weight maps which needs to be taken into account.  In order to retrieve a signal-to-noise ratio standard deviation of unity, and a null mean, assuming a Gaussian noise distribution, we perform a Gaussian fitting on the S/N pixel histogram values between -3 and 1.5 to avoid the contamination by the sources. The best-fit parameters provide the global offset of the background and scale of the matched filter S/N pixel values, which are measured by the center ($\mu_{S/N}$) and the width ($\sigma_{S/N}$) of the best-fit Gaussian function. Slight variations from unity are expected in case of residual correlated noise in the maps or small bias in the absolute background value. We thus normalize the noise maps by $\sigma_{S/N}$ to have a unity standard deviation in the S/N map : 
\begin{equation}
\label{eq:check_snr}
N_{corr} = N_{ori} \times \sigma_{S/N},
\end{equation}
where $N_{corr}$ is the noise map after this correction and $N_{ori}$ is the original noise map. Similarly, we correct the S/N maps using
\begin{equation}
\label{eq:check_snr2}
        S/N_{corr} = \frac{S/N_{ori} - \mu_{S/N}}{\sigma_{S/N}},
\end{equation}
where $S/N_{corr}$ is the corrected S/N map and $S/N_{ori}$ is the original S/N map.\\

In absence of noise, a matched-filter S/N map of an isolated point source is maximal at the position of the source. The \textit{nikamap} package incorporates the \textit{find\_peak} algorithm from the \textit{photutils} package \citep{Bradley+22}, to identify peaks above a certain threshold in the matched-filter S/N maps. 
We set an extraction box size to 5 pixels, allowing only the brightest source in a 2.5 pixel distance on both axes to be extracted. The box size is chosen to be matched with the FWHM of the PSF. The value at the position of detection in the beam matched-filter S/N maps provide the S/N of the point sources. 
We finally perform PSF-fitting photometry on the original PIIC signal maps (which are shown in the Appendix\,\ref{sc:N2CLS maps}) with the \textit{nikamap} package based on the \textit{BasicPSFPhotometry} module in \textit{photutils}. The position of the sources is fixed and corresponds to the results of \textit{find\_peak}. The PSFs used in this process are two-dimensional circular Gaussians that have their FWHMs defined by the beam described in the PIIC data products. The backgrounds of the maps are estimated by the \textit{MedianBackground} module in \textit{photutils} and removed during the analysis.\\
For the source detection in the N2CLS observations, we further refine the choice of detection threshold in both fields and bands according to the purity analysis presented in Sect.~\ref{sc:purity}. We use the source fluxes from the PSF fitting in the following analysis, which provide more robust flux measurements on slightly blended sources than the peak flux. As for the flux uncertainties, we provide the pixel value of the noise map after corrections (Eq.~\ref{eq:check_snr}), which accounts for the noise correlations between nearby pixels. It does not take into account the additional uncertainties caused by degeneracies between the individual fluxes of heavily blended galaxies. However, since we do not deblend sources closer than 2.5 pixels, this is not a problem.

\section{Characterization of the source extraction performances}\label{sc:flux_bias_corr}

\subsection{The simulation framework}\label{sc:simulation_framework}

In far-IR and millimeter blind surveys, flux measurements on individual sources and source number counts estimates are affected by systematic effects from data reduction, source detection, and flux measurement. Previous studies have developed various methods to estimate and correct these effects. These include Bayesian techniques measuring the posterior distribution of source fluxes detected above certain S/N under a given source number counts, which was applied in early single-dish surveys \citep{Coppin+05}. Some more recent studies turn to more empirical methods. They generate a series of pure-noise half-difference maps produced by randomly inverting the signal in half of the data/sub-scans and inject sources in them following a given galaxy number counts. Then they repeat the source detection and flux measurement procedure and estimate the deviation of the output flux from the input flux versus other properties and apply the correction to the real source catalog \citep[e.g.][]{Geach+17,Liu2018,Zavala+21}. Empirical methods are also applied to estimate the false detection rate and the completeness at a given flux or S/N, which are then applied to the estimate of source number counts.\\

The Bayesian technique and empirical methods in previous studies mainly account for the impact on source detection and flux measurement of instrument noise. In addition to instrument noise, previous studies also pointed out that astrophysical clustering and random alignments of sources inside the beam could also have a non-negligible impact on detected source fluxes in the $\sim$3m (far-infrared), $\sim$15m (sub-millimeter) and $\sim$30m (millimeter) single-dish observations \citep{Bethermin+17}.\\

In addition, we also need to quantify the impact on the source fluxes of filtering that results from the data reduction pipeline, i.e. transfer functions. They have been first measured for NIKA by \citet{Adam2015} and further explored for the reduction of NIKA2 observations of the Sunyaev-Zeldovich effect in galaxy clusters \citep[e.g.][]{Munoz2023}. Our N2CLS observations, aiming to detect faint point sources in deep field, use different data reduction methods and we thus need to measure our own transfer function.\\

PIIC offers the possibility to inject artificial sources, or an artificial sky map into the timelines (since its "May21" version), allowing us to take into account the impact from both instrument noise and pipeline artifacts. A beam-convolved, noiseless sky model in the corresponding band is used as an input. The sky model is then translated into timelines of individual NIKA2 KIDs. These noiseless time series from each KID are then combined with data from real calibrated observations with sign flipped every other map. This process generates timeline data that resemble NIKA2 observations of the modeled sky region with the same depth as real observations but free from real astrophysical signal. These data are then reduced by the normal PIIC data reduction pipeline to produce simulated N2CLS-like maps. In the PIIC reduction and map construction with simulated timelines, we use the same parameters as for the reduction of N2CLS observations (see Sect.~\ref{sc:data_red}). \\

The input sky model is provided by the simulated infrared dusty extragalactic sky \citep[SIDES,][]{Bethermin+17,Bethermin+22} simulation. We use the SIDES light cone based on the Uchuu dark-matter simulation \citep{Ishiyama+21} presented in \citet{Gkogkou+22}. For each dark-matter halo, galaxy properties are generated following empirically measured scaling relations. The 1.2 and 2\,mm fluxes are then derived using the NIKA2 bandpass, and maps are produced based on the fluxes and positions of all galaxies. The maps are then smoothed to the resolution of NIKA2 at the corresponding wavelength, as required by the PIIC simulation. Using a dark matter simulation, we obtain simulated sky maps with realistic galaxy clustering between sources, which is not accounted for in the techniques used in most of the previous studies. From the 117\,deg$^2$ simulation, a total of 117 independent tiles of 1\,deg$^2$ are used to produce the input sky model at 1.2 and 2\,mm simulating the GOODS-N and COSMOS fields, which later produce 117 independent simulated N2CLS observations at the two wavelengths on the two fields.
For each simulation, the scans list were shuffled before being read by PIIC. This ensures the noise realisations to be independent in the 117 simulations of each field. \\

From the 117 input sky models for each field and band, we also obtain two sets of catalogs. The "galaxy catalogs" record the fluxes assigned to each simulated galaxy, blended or not with nearby galaxies in the NIKA2 beam. From the noiseless beam matched-filter model map, we also identify all peaks above a certain peak flux and record their position and peak fluxes in the "blob catalogs". The peak flux thresholds are set to 0.15/0.05\,mJy for GOODS-N and 0.45/0.15 \,mJy for COSMOS for input maps of simulation at 1.2/2\,mm. These thresholds are comparable to the noise levels in the corresponding NIKA2 maps (see Table\,\ref{tab:tab-surveys}), and are below the detection limits in the noisy simulated maps. In the analysis of completeness, purity and flux correction, we will use the input "blob catalog" as the reference, which is subject to the same source blending at the same NIKA2 angular resolution. The galaxy catalog will be used to correct the impact of source blending on the number counts, as described in Sect.~\ref{sc:numbercounts_source_to_gal_corr}.

   \begin{figure*}[tb]
   \centering
      \includegraphics[clip=true,width=.99\textwidth]{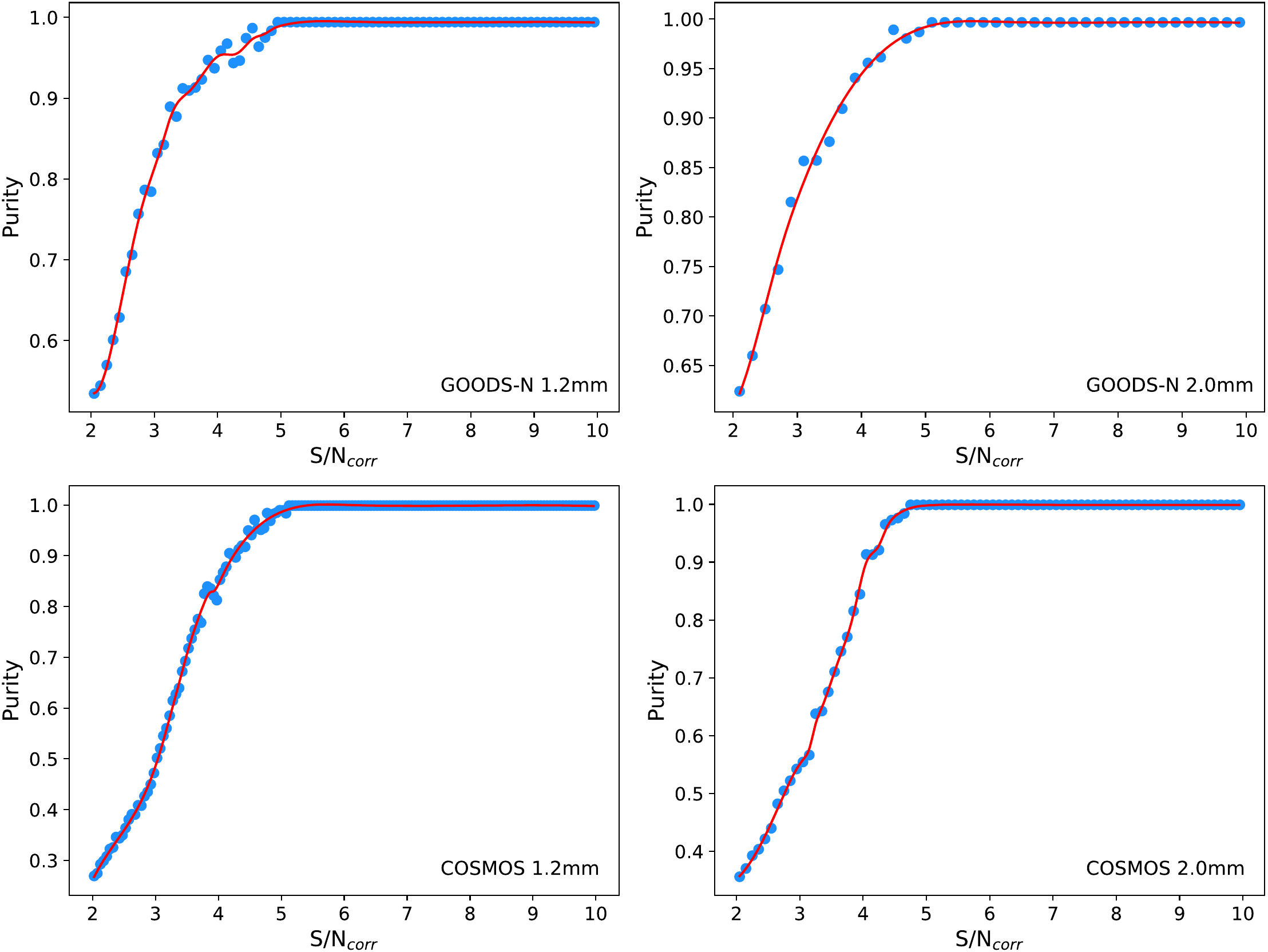}
      \caption{Purity of detected sources at different S/N in the matched-filter map from the simulations at 1.2 and 2\,mm in the GOODS-N (top panels) and COSMOS (lower panels) field. }
      \label{fig:purity}
   \end{figure*}

\subsection{Purity of detection}\label{sc:purity}

We first use the 117 simulated observations to determine the purity of the extracted source candidate sample. The purity is defined as the probability for a source extracted from the simulated output map to be real. In practice, we consider that an extracted source is real if it can be matched with a source from the input blob catalog. The matching radius in position is 0.75 times the FWHM of the Gaussian beam of each NIKA2 band, which is consistent with the distance threshold used in previous studies \citep{Geach+17}. The same threshold of source cross-matching is also used to estimate the source completeness (see Sect.~\ref{sc:completeness}), which ensures the two effects are consistently accounted for in the following estimate of source number counts. As described in Sect.~\ref{sc:simulation_framework} we limit the cross-matching to sources above certain fluxes. In cases of multiple catalog sources falling within the matching radius, we recognize the brightest source as the counterpart in the following analysis.\\

We present the purity as a function of S/N in Fig.~\ref{fig:purity}. The results are fitted by a spline function. In COSMOS, the purity at 1.2 and 2\,mm reaches 80\% at S/N$_{corr}$=3.8, and is $>$95\% at S/N$_{corr}>$4.6. In GOODS-N, the purity reaches 80\% at S/N$_{corr}$=3.0 and S/N$_{corr}$=2.9, at 1.2 and 2\,mm respectively, and reaches $>$95\% at S/N$_{corr}>$4.2 and S/N$_{corr}>$4.1. The noise level is lower in GOODS-N than in COSMOS. The source density is thus higher in GOODS-N at fixed S/N threshold, while the number density of spurious sources is similar for the same S/N threshold. A given purity is thus reached for a lower S/N threshold in GOODS-N than in COSMOS. This analysis allows us to set the S/N detection threshold to reach a 80\% purity in each field and wavelength. For the number counts (Sect.\,\ref{sc:numbercounts_calc}), the contamination by spurious sources can be corrected statistically and we thus choose a 80\,\% purity, such as to have corrections of 20\,\% at most. In contrast, a $>$95\,\% purity could be considered to select sources for follow-up observations.

\begin{figure*}[tb]
   \centering
      \includegraphics[clip=true,width=.99\textwidth]{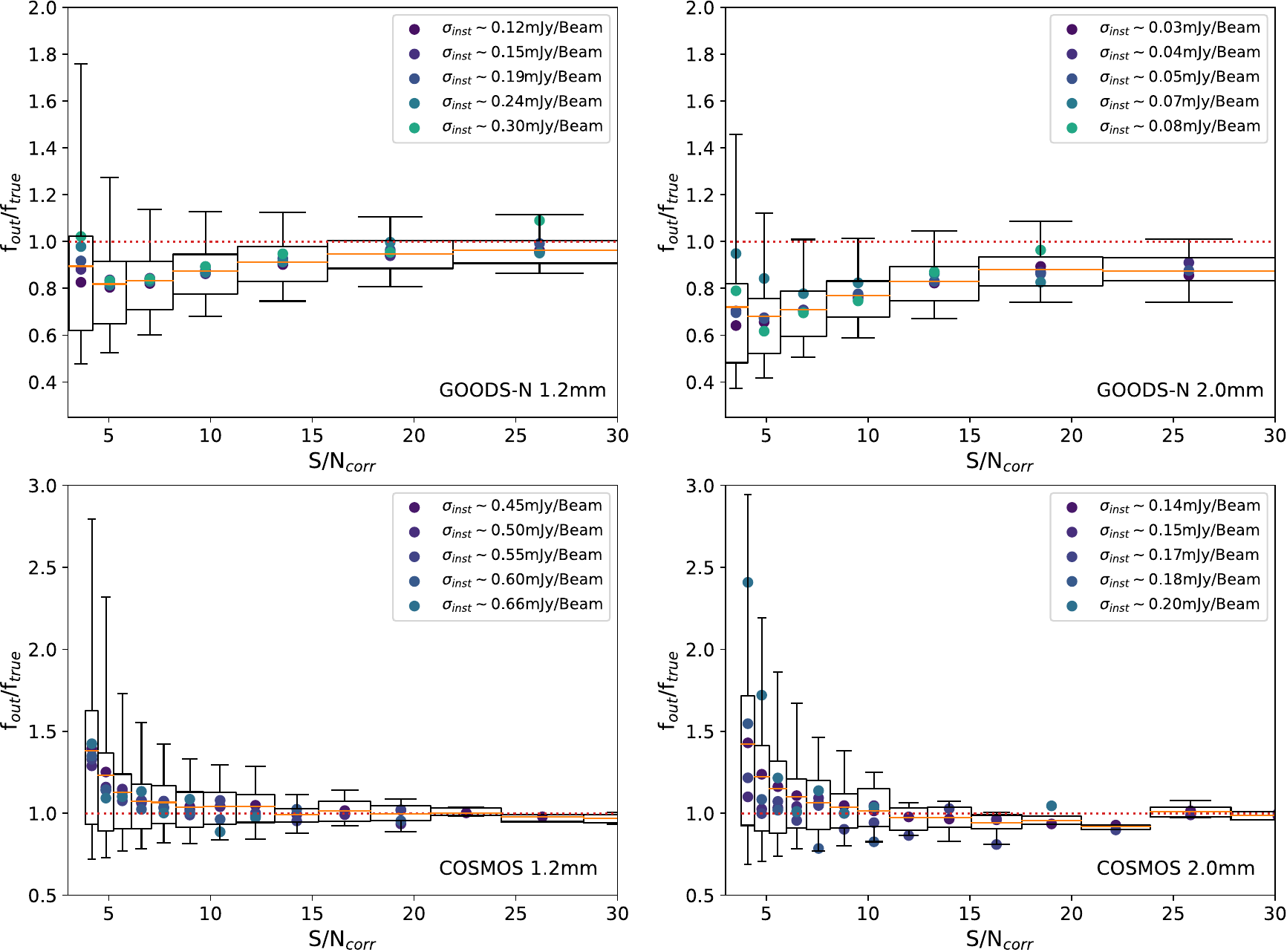}
      \caption{Ratio between the source fluxes measured in the output simulated maps ($f_{out}$) and the source fluxes from the input blob catalog ($f_{true}$) as functions of S/N at 1.2\,mm (left panels) and 2\,mm (right panels) in GOODS-N (upper panels) and COSMOS (lower panels). This corresponds to the effective flux boosting described in Sect.\,\ref{sc:flux_boosting}. The boxes shown for each S/N bin represent ranges between 25\% to 75\% of the cumulative distribution and the upper and lower bounds of error bars present the 5\% to 95\% of the cumulative distribution (within 1$\sigma$). The width of each box corresponds to the width of the corresponding S/N$_{corr}$ bin. The red dotted line shows the positions of unity effective flux boosting as a reference of each plot. In addition, we use color coded solid filled circles to present the median flux boosting in regions with different noise levels. }
      \label{fig:fboost_eff}
   \end{figure*}

\subsection{Effective flux boosting} \label{sc:flux_boosting}

We also have to evaluate the impact of noise and data reduction (pipeline transfer function) on the measured source fluxes. We estimate these effects by comparing the recovered flux to the input one for each blob of our simulation. \\

Like most blind single-dish deep-field surveys in the far-IR and millimeter, our blind detection algorithm uses only S/N as a threshold parameter. Considering the existence of noise in the real data, this method naturally biases detections towards sources that coincide with noise peaks. This boosts faint source fluxes above the threshold and leads to systematically overestimated fluxes for these objects. This effect is called "flux boosting".\\

Apart from the flux-boosting effect, the PIIC pipeline could filter out a fraction of the source flux density. As explained in Sect.\,\ref{sc:data_red}, PIIC is using the most correlated KIDS to estimate and remove sky noise per KID. An additional polynomial baseline is removed per sub-scan to correct for remaining instabilities. Due to the iterative mode of the data reduction (which is based on a S/N threshold to build a "source map"), sources at lower S/N are more affected than sources at higher S/N by filtering effects.\\

A proper correction of the combination of all the effects on source flux measurements is crucial to estimate source number counts. However, the detailed analysis of the contribution of each individual effect is beyond the scope of our work. In this paper, we directly measure the effective ratios between flux densities measured in the simulated maps and those in the input blob catalogs, and study the variation of the ratios with S/N$_{corr}$. Both the flux boosting and the pipeline filtering contribute to the deviation between input and output fluxes, and we call this output over input flux ratio as effective flux boosting. To estimate the effective flux boosting of source detection for each field and at each wavelength, we cross-match the input blob catalogs to the sources detected by our extraction algorithm in the output simulated maps. An input blob is considered to be recovered by our detection algorithm if any source above the S/N threshold (see Sect.~\ref{sc:purity}) could be found within 0.75$\times$FWHM to its position. If multiple input blobs correspond to the same detection, we use the closest one. \\

 The distributions of the effective flux boosting are presented in Fig.~\ref{fig:fboost_eff}. We remark that the variation of median effective flux boosting between regions with different noise levels are small compared to the scatter. We thus only focus on the variation of effective flux boosting versus S/N$_{corr}$ in the following analysis and discussion. 
 
 The mean effective flux boosting for both bands in the COSMOS field is mostly above one, suggesting that it is mainly dominated by the flux boosting due to instrument noise. The mean effective flux boosting curve also well matches the functional correction used by the S2CLS survey \citep{Geach+17}, which only accounts for the typical flux boosting in their simulation. Contrary to COSMOS, the mean effective flux boosting in GOODS-N field drops below unity even at relatively high S/N and reaches $\sim$0.8 at the S/N used as detection limit in both bands. This suggests a much stronger filtering effect on source flux densities by the data reduction pipeline. \\
 
In Fig.~\ref{fig:fboost_eff}, we also notice a significant scatter in the ratio between input and output fluxes, especially at low S/N. The interquartile range is as high as 40-80\% at the S/N$_{corr}\sim$4, and drops to $\sim$10\% or less at S/N$_{corr}>$20. Even if we know the average correction to apply as a function of S/N$_{corr}$, there are large uncertainties on this correction at low S/N$_{corr}$. In Sect.\,\ref{sc:numbercounts_calc}, we discuss how this is taken into account to derive the number counts.

\begin{figure*}[tb]
   \centering
      \includegraphics[clip=true,width=.99\textwidth]{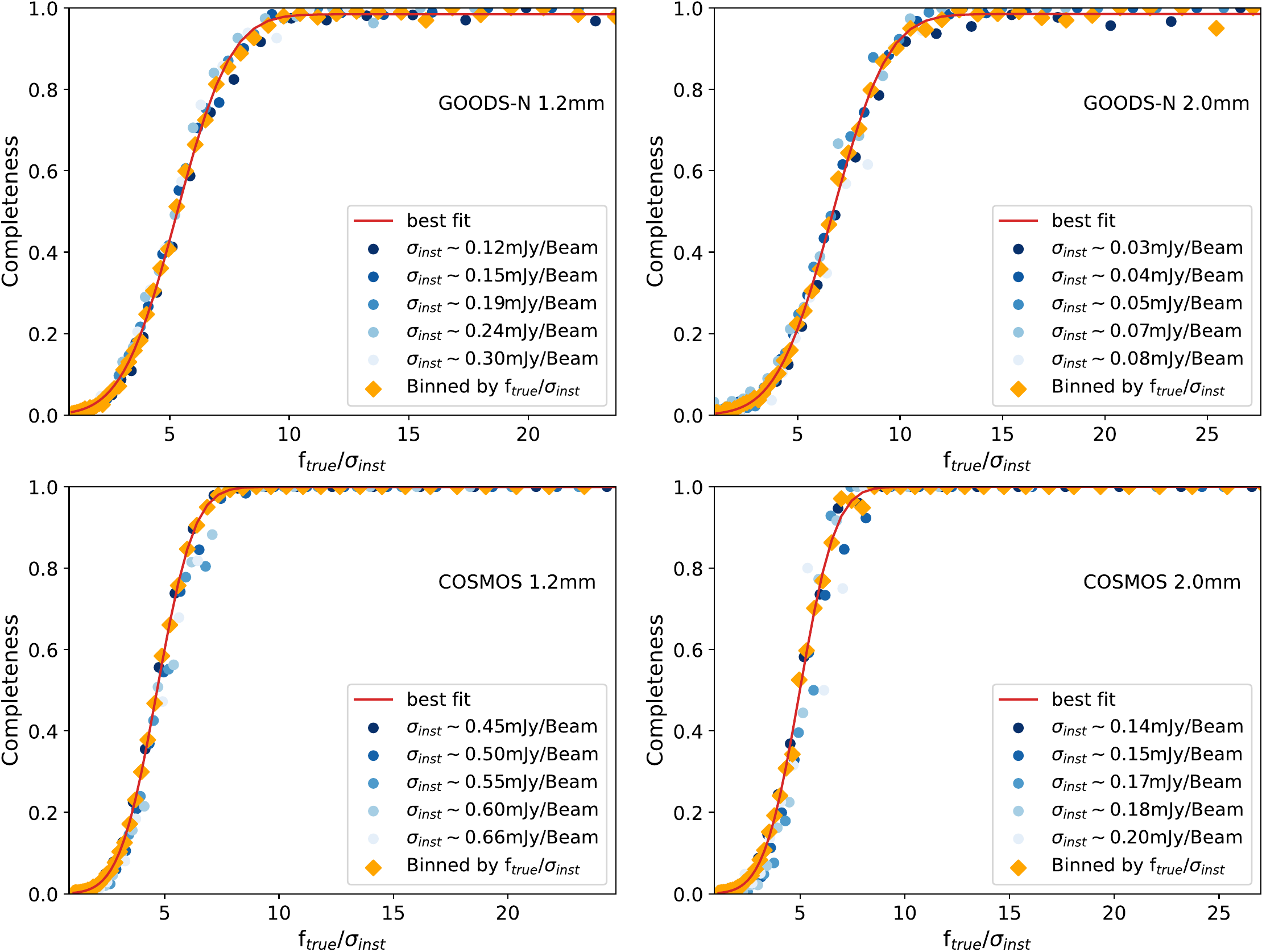}
      \caption{Completeness of sources in N2CLS as a function of the ratio between source flux density and instrumental noise level (f$_{true}$/$\sigma_{inst}$) at 1.2\,mm (left panels) and 2\,mm (right panels) in GOODS-N (upper panels) and COSMOS (lower panels). The completeness in various survey areas with different instrument noise level are presented as bluish color coded dots, and the average completeness over whole survey area is shown using oranges dots. The red line shows the best fit of the average completeness using the functional form of Eq.~\ref{eq:comp_model}. }
      \label{fig:completeness}
   \end{figure*}

\subsection{Completeness}\label{sc:completeness}

Another key information to derive the number counts from our survey is the completeness of our catalogs. The completeness is defined as the probability to find a source in the output catalog as a function of its input properties. For an heterogeneous survey, it can also vary depending on the position of the source. Since our sources are unresolved, we consider only the input blob flux density (f$_{true}$) as a relevant physical parameter. Concerning the variable linked to the survey, it is mainly driven by the instrument noise level at the source position ($\sigma_{inst}$).\\

In practice, the completeness is evaluated as the fraction of input blobs at a given flux density (f$_{true}$) that are recovered in the output catalog. The completeness curve varies significantly if we compute it in regions with different noise levels. However, by computing the completeness as a function of f$_{true}$/$\sigma_{inst}$, i.e. the input flux divided by the noise level at the source position, we find a similar completeness function in all regions, as shown in Fig.\,\ref{fig:completeness}. This highlights that the completeness is a function of two main parameters: f$_{true}$ and $\sigma_{inst}$. We fit our results using the following functional form:
\begin{equation}
\label{eq:comp_model}
\mathcal{C}(x) = \frac{1+\mathrm{erf} \left ( \frac{x-x_0}{y_0} \right )}{2},
\end{equation}
where $\mathcal{C}$ is the completeness, and $x_0$ and $y_0$ are free parameters. In our case, $x$ is f$_{true}$/$\sigma_{inst}$. These best fits are used further in the paper to derive the completeness of sources at given fluxes and compute their corresponding effective area of the survey (Sect.\,\ref{sc:eff_area}).\\

With the purity, effective flux boosting and completeness correction functions set, here we summarize the properties of N2CLS sample from blind detection for the following analysis in the paper. With the S/N thresholds of 80\% purity set in Sect.~\ref{sc:purity}, we detect 120 and 195 sources at 1.2\,mm in GOODS-N and COSMOS respectively, and 67 and 76 sources at 2\,mm. In the 1.2\,mm maps of GOODS-N and COSMOS, we detect sources as faint as 0.4\,mJy and 1.7\,mJy in uncorrected PSF fluxes. At 2\,mm, we reach limiting uncorrected PSF fluxes of 0.1\,mJy and 0.5\,mJy in GOODS-N and COSMOS, respectively. \\

The GOODS-N sample includes 44 sources with both 1.2\,mm and 2.0\,mm detection, 76 sources with only 1.2\,mm detection and 23 sources with only 2.0\,mm detection. The COSMOS sample includes 49 sources with both 1.2\,mm and 2.0\,mm detection, 146 sources with only 1.2\,mm detection and 27 sources with only 2.0\,mm detection. This large number of sources detected only at 2\,mm could seem surprising considering the model forecasts, and is discussed in Sect.\,\ref{sect:2mmonly_model}.

\subsection{Effective survey area}\label{sc:eff_area}

To derive the surface density of sources (also called number counts, see Sect.~\ref{sc:numbercounts}), we need to know the surface area of our survey. However, the survey has no clear border, since faint sources are unlikely to be detected in the noisy outskirts of the field. To take this into account, we use a similar method as \citet{Bethermin+20}. For a given source flux density, each pixel of the map is associated to a different completeness. The effective area is then the sum of the surface area of each pixel ($\Omega_{\mathrm{pix}}$) multiplied by the completeness at the pixel position,
\begin{equation}
\label{eq:effarea}
\Omega_{\mathrm{eff}}(S) = \sum_{i=1}^{N_{\rm pixel}}\Omega_{\mathrm{pix}} \times \mathcal{C}\left(\frac{S}{\sigma_{i}}\right),
\end{equation}
where $\mathcal{C}\left(\frac{S}{\sigma_{i}}\right)$ is the completeness expected for an hypothetical source with a true flux density $S$. In practice, $S$ is not known and we use the deboosted flux density, which is the raw flux corrected by the expected effective flux boosting factor at the S/N of detection, as a proxy (see Sect.\,\ref{sc:flux_boosting}). The $\sigma_{i}$ value is the instrument noise level in the i-th pixel. This quantity varies with the source flux density, since fainter sources are less likely to be detected in the noisy edge of the map. Sources with lower flux densities will thus be associated to smaller effective areas.\\

This computation naturally accounts for the non-uniform depth within the N2CLS maps. It is especially crucial for our number counts analysis in GOODS-N field with 3 times variation in survey depth across the region considered for number counts analysis.

    \begin{table*}
    \caption{\label{tab:1.2_source} Differential and cumulative source number counts at 1.2\,mm in GOODS-N and COSMOS. The columns S and S$_\mathrm{min}$ refer to the midpoint and the lower bound flux of the flux bins used to compute the number counts. The differential and cumulative number counts are dN/dS and N(>S). R$_\mathrm{diff}$ and R$_\mathrm{cum}$ represent the source-to-galaxy counts correction factor (see Sect.\,\ref{sc:numbercounts_source_to_gal_corr}). The average number of sources within each differential flux bin among the 100 Monte Carlo realizations are presented in column <N> of each field and array (see Sect\,\ref{sc:numbercounts_calc}).}
        \begin{tabular}{ccccccclccccccc}\toprule
            \multicolumn{7}{c}{\textbf{GOODS-N}}& &\multicolumn{7}{c}{\textbf{COSMOS}}
            \\\cmidrule(r){1-7}\cmidrule(r){9-15}
            S & <N> & dN/dS& R$_\mathrm{diff}$ & S$_\mathrm{min}$ & N($\mathrm{>}$S) & R$_\mathrm{cum}$ & & S & <N> & dN/dS& R$_\mathrm{diff}$ & S$_\mathrm{min}$ & N($\mathrm{>}$S) & R$_\mathrm{cum}$\\
            mJy &  & deg$^\mathrm{-2}$mJy$^\mathrm{-1}$ & & mJy & deg$^\mathrm{-2}$ & & & mJy & & deg$^\mathrm{-2}$mJy$^\mathrm{-1}$ & & mJy &  deg$^\mathrm{-2}$ & \\\midrule
        0.66 & 10.9 & 6799$^\mathrm{+3398}_\mathrm{-2938}$ & 0.76 & 0.60 & 3598$^\mathrm{+448}_\mathrm{-534}$ & 0.71 &  & 1.75 & 24.2 &  1041$^\mathrm{+312}_\mathrm{-272}$ & 0.68 & 1.57 & 948$^\mathrm{+108}_\mathrm{-102}$ & 0.70 \\
        0.81 & 13.6 & 4650$^\mathrm{+1918}_\mathrm{-1518}$ & 0.72 & 0.73 & 2727$^\mathrm{+345}_\mathrm{-389}$ & 0.69 &  & 2.12 & 31.3 & 592$^\mathrm{+155}_\mathrm{-140}$ & 0.71 & 1.92 & 597$^\mathrm{+69}_\mathrm{-69}$ & 0.72 \\
        0.98 & 14.6 & 2960$^\mathrm{+1157}_\mathrm{-981}$ & 0.71 & 0.88 & 2005$^\mathrm{+276}_\mathrm{-311}$ & 0.68 &  & 2.57 & 31.4 & 315$^\mathrm{+85}_\mathrm{-76}$ & 0.72 & 2.32 & 355$^\mathrm{+48}_\mathrm{-48}$ & 0.72 \\
        1.19 & 13.1 & 1748$^\mathrm{+747}_\mathrm{-630}$ & 0.66 & 1.07 & 1447$^\mathrm{+229}_\mathrm{-286}$ & 0.66 &  & 3.12 & 24.3 & 159$^\mathrm{+47}_\mathrm{-40}$ & 0.73 & 2.82 & 199$^\mathrm{+34}_\mathrm{-31}$ & 0.72 \\
        1.44 & 11.4 & 1089$^\mathrm{+495}_\mathrm{-446}$ & 0.64 & 1.30 & 1047$^\mathrm{+196}_\mathrm{-243}$ & 0.66 &  & 3.79 & 15.2 & 75.8$^\mathrm{+28.7}_\mathrm{-25.0}$ & 0.69 & 3.42 & 103$^\mathrm{+25}_\mathrm{-25}$ & 0.71 \\
        1.75 & 8.8 & 627$^\mathrm{+337}_\mathrm{-248}$ & 0.66 & 1.58 & 745$^\mathrm{+176}_\mathrm{-200}$ & 0.67 &  & 4.59 & 5.6 & 22.5$^\mathrm{+17.0}_\mathrm{-13.9}$ & 0.72 & 4.15 & 47.7$^\mathrm{+18.2}_\mathrm{-15.6}$ & 0.74 \\
        2.12 & 8.9 & 501$^\mathrm{+264}_\mathrm{-229}$ & 0.72 & 1.92 & 533$^\mathrm{+160}_\mathrm{-187}$ & 0.69 &  & 5.58 & 2.5 & 8.41$^\mathrm{+10.45}_\mathrm{-7.09}$ & 0.72 & 5.04 & 27.8$^\mathrm{+15.1}_\mathrm{-11.9}$ & 0.76 \\
        2.57 & 6.9 & 315$^\mathrm{+198}_\mathrm{-145}$ & 0.66 & 2.32 & 328$^\mathrm{+148}_\mathrm{-177}$ & 0.66 &  & 6.76 & 3.4 & 9.18$^\mathrm{+8.41}_\mathrm{-6.03}$ & 0.71 & 6.11 & 18.8$^\mathrm{+13.2}_\mathrm{-9.4}$ & 0.83 \\
        3.12 & 3.5 & 130$^\mathrm{+128}_\mathrm{-86}$ & 0.67 & 2.82 & 172$^\mathrm{+162}_\mathrm{-154}$ & 0.66 &  & 8.21 & 1.5 & 3.35$^\mathrm{+5.69}_\mathrm{-2.67}$ & 1.19 & 7.42 & 6.81$^\mathrm{+11.31}_\mathrm{-5.90}$ & 0.95 \\

            \bottomrule
        \end{tabular}
    \end{table*} 
    
    \begin{table*}
    \caption{\label{tab:2.0_source}Similar to Table.~\ref{tab:1.2_source} but at 2\,mm.}
        \begin{tabular}{ccccccclccccccc}\toprule
            \multicolumn{7}{c}{\textbf{GOODS-N}}& &\multicolumn{7}{c}{\textbf{COSMOS}}
            \\\cmidrule(r){1-7}\cmidrule(r){9-15}
            S & <N> & dN/dS& R$_\mathrm{diff}$ & S$_\mathrm{min}$ & N($\mathrm{>}$S) & R$_\mathrm{cum}$ & & S & <N> & dN/dS& R$_\mathrm{diff}$ & S$_\mathrm{min}$ & N($\mathrm{>}$S) & R$_\mathrm{cum}$\\
            mJy &  & deg$^\mathrm{-2}$mJy$^\mathrm{-1}$ & & mJy & deg$^\mathrm{-2}$ & & & mJy & & deg$^\mathrm{-2}$mJy$^\mathrm{-1}$ & & mJy  & deg$^\mathrm{-2}$ & \\\midrule
           0.24 & 11.2 & 12340$^\mathrm{+5732}_\mathrm{-5009}$ & 0.59 & 0.20 & 2073$^\mathrm{+378}_\mathrm{-385}$ & 0.55 &  & 0.61 & 17.7 & 1186$^\mathrm{+438}_\mathrm{-352}$ & 0.49 & 0.52 & 358$^\mathrm{+73}_\mathrm{-70}$ & 0.51 \\
           0.33 & 12.2 & 4981$^\mathrm{+2243}_\mathrm{-1951}$ & 0.53 & 0.27 & 1157$^\mathrm{+237}_\mathrm{-240}$ & 0.51 &  & 0.84 & 15.5 & 314$^\mathrm{+124}_\mathrm{-107}$ & 0.51 & 0.71 & 131$^\mathrm{+33}_\mathrm{-32}$ & 0.53 \\
           0.45 & 11.8 & 2428$^\mathrm{+1140}_\mathrm{-880}$ & 0.51 & 0.38 & 650$^\mathrm{+168}_\mathrm{-182}$ & 0.49 &  & 1.14 & 7.7 & 80.1$^\mathrm{+43.4}_\mathrm{-37.6}$ & 0.53 & 0.97 & 48.4$^\mathrm{+18.9}_\mathrm{-16.1}$ & 0.58 \\
           0.61 & 8.5 & 1077$^\mathrm{+590}_\mathrm{-473}$ & 0.47 & 0.52 & 311$^\mathrm{+122}_\mathrm{-118}$ & 0.47 &  & 1.57 & 3.7 & 27.0$^\mathrm{+24.1}_\mathrm{-18.3}$ & 0.68 & 1.33 & 19.6$^\mathrm{+14.0}_\mathrm{-9.9}$ & 0.73 \\
           0.84 & 2.9 & 249$^\mathrm{+308}_\mathrm{-212}$ & 0.43 & 0.71 & 104$^\mathrm{+105}_\mathrm{-100}$ & 0.46 &  & 2.16 & 1.4 & 7.50$^\mathrm{+13.21}_\mathrm{-5.96}$ & 0.70 & 1.82 & 6.28$^\mathrm{+10.82}_\mathrm{-5.37}$ & 0.82 \\
            \bottomrule
        \end{tabular}
    \end{table*}

\section{Number Counts}\label{sc:numbercounts}

\subsection{Derivation of the source number counts}\label{sc:numbercounts_calc}

The surface density of sources per observed flux density interval, or differential source number counts (dN/dS), is a simple measure of redshift-integrated source abundance. It has been used as a powerful tool to test and compare galaxy evolution models \citep[e.g][]{McAlpine+19,Lagos+20,Lovell+21}. We derive the differential source number counts at 1.2 and 2\,mm in a given flux density bin using:
\begin{equation}
\label{eq:dnds}
\frac{dN(S)}{dS} = \frac{1}{\Delta S} \sum_{j=1}^{N_{\rm source}}\frac{\mathcal{P}_j}{\Omega_{\mathrm{eff},j}(S_j)},
\end{equation}
where $\mathcal{P}_j$ and $\Omega_{\mathrm{eff},j}$ are the purity and effective survey area associated to the j-th source of the extracted catalog and with a flux density inside the bin (see Sect.~\ref{sc:purity} and Sect.~\ref{sc:eff_area}, respectively), and $\Delta S$ is the width of the flux bin.\\

We also derive the corresponding cumulative source number counts (N(>S)). They are defined as the surface density of sources above a certain flux density higher than a given value, and estimated using
\begin{equation}
\label{eq:ngts}
N(>S) = \sum_{k=1}^{n}\frac{\mathcal{P}_k}{\Omega_{\mathrm{eff},k}(S_k)}\,.
\end{equation} 
Contrary to the differential number counts, we sum all the sources with a flux density above a certain threshold instead of only the ones in the bin.\\

As discussed in Sect.\,\ref{sc:eff_area}, we use the deboosted flux density as a proxy for the true flux density in the computation of the effective survey area. To take into account the effect of the uncertainties on the deboosting factors, we perform 100 Monte Carlo realizations in which we draw a deboosting factor from the distribution and derive the deboosted flux density for each source, and finally compute the number counts accordingly. The median of the number counts realization is used to determine the central value and the 16th and 84th percentiles used to compute the uncertainties. Finally, we combine quadratically these uncertainties from our Monte Carlo method with Poisson uncertainties.\\

To derive the differential number counts, we choose a common flux density binning for GOODS-N and COSMOS to allow later on an easier combination of the two measurements. The lower bound of the faintest bin is defined to have at least 50\,\% completeness in the deepest field (GOODS-N). This corresponds to 0.60 and 0.20\,mJy at 1.2 and 2\,mm, respectively. The upper bound of the brightest bin is set to a slightly higher value than the brightest source of the survey. This corresponds to a flux density of 9.0 and 2.5\,mJy at 1.2 and 2\,mm, respectively. We use an uniform logarithmic sampling of this range. The number of bins is a compromise between a good sampling in flux density and a sufficient number of source per bin to have reasonable uncertainties. We use 14 and 8 bins at 1.2 and 2\,mm, respectively. In the faintest bins (5 bins at 1.2\,mm and 3 bins at 2\,mm), the completeness in the wider but shallower COSMOS field is well below 50\,\%. Large and unreliable corrections are thus necessary, and we thus do not compute the number counts in this regime for this field. In the brightest bins (5 bins at 1.2\,mm and 3 bins at 2\,mm), GOODS-N is too small to contain any source and number counts cannot be derived.\\

The cumulative numbers counts are derived using a similar method. The lower bounds of the differential number count bins are used as flux density limits. Our results are summarized in Table\,\ref{tab:1.2_source} and Table\,\ref{tab:2.0_source}.

   \begin{figure*}[tb]
   \centering
      \includegraphics[clip=true,width=.99\textwidth]{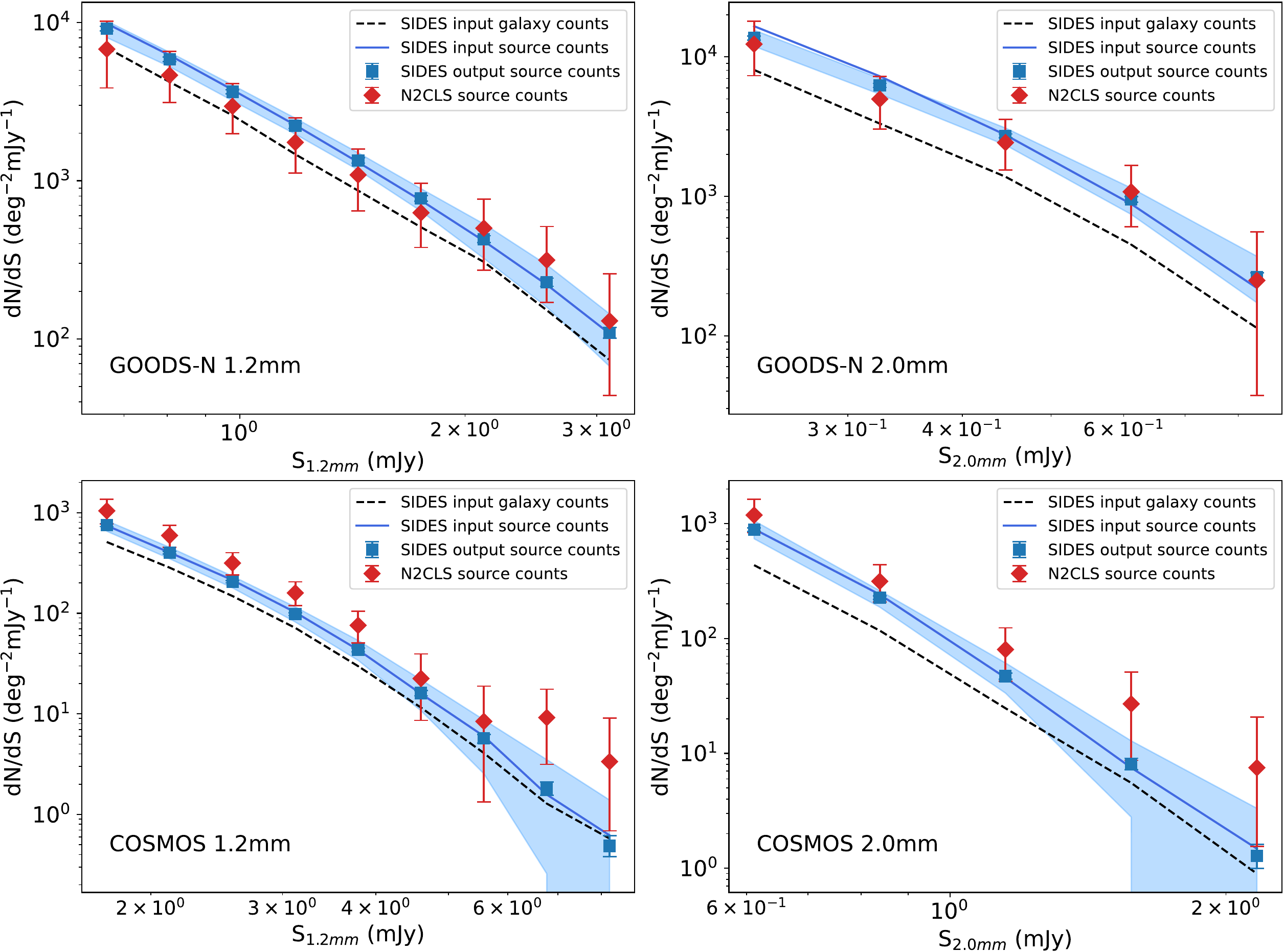}
      \caption{\label{fig:num_simu_in_vs_out} Comparison between the differential number counts from simulations and observations at 1.2\,mm (left panels) and 2\,mm (right panels) in GOODS-N (upper panels) and COSMOS (lower panels). The solid blue line represents the source number counts derived directly from SIDES noiseless maps (see Sect.\,\ref{sc:numbercounts_calc_validation}), while the blue squares are the source number counts estimated using the full analysis pipeline (map making, source extraction, and statistical corrections). They are essentially the input and output source number counts from the simulation. The blue shaded area illustrates the 1$\sigma$ field-to-field variance of the output source number counts. Since several galaxies can contribute to a source, we also show the galaxy number counts in SIDES as black dashed line (see discussion in Sect.\,\ref{sc:numbercounts_source_to_gal_corr}). Finally, the number counts measured from N2CLS are represented by red diamonds.}
   \end{figure*}

\subsection{Validation of the number counts reconstruction from simulations}\label{sc:numbercounts_calc_validation}

Before comparing our number counts with previous observations, we perform an end-to-end simulation of the full analysis process to validate its accuracy. We thus apply the same exact algorithm to derive the number counts from our 117 simulated fields based on SIDES (see Sect.\,\ref{sc:flux_bias_corr}). These simulations will thus include all the possible instrumental effects, such as the transfer function of the map making or the source extraction biases. We derive the source number counts by combining the 117 simulated fields to obtain output number counts with low uncertainties. We also derive the source counts from each individual simulated field to derive the field-to-field variance (also referred as cosmic variance in the literature).\\

In Fig.\,\ref{fig:num_simu_in_vs_out}, we compare these output number counts from our end-to-end simulation (blue squares) with the input source number counts derived from the blob catalog extracted from the noiseless SIDES maps (blue solid line, see Sect.\,\ref{sc:simulation_framework} for the description of blob catalogs). The input and output results agree at 1\,$\sigma$, except mild ($<$20\,\%) disagreement in the faintest GOODS-N bin at 2\,mm. This demonstrates the robustness of our method to derive source number counts.\\

We also compared the source number counts from our simulation and the N2CLS data (red diamonds). They always agree at better than 2\,$\sigma$ and the majority of the data points are in the 1\,$\sigma$ range. The SIDES simulation is thus very close from the real observations. This justifies a posteriori the choice of this simulation to characterize the map making and the source extraction effects.\\
We can see however that there is a small systematic offset between the simulation and the observations with GOODS-N being lower than SIDES and COSMOS being higher. This could be caused by the field-to-field variance, since flux bins are usually correlated (see e.g. \citet{Gkogkou+22} for the case of line-luminosity functions). We derived the 1\,$\sigma$ range of the number counts from the 117 SIDES realizations and found that the offset between N2CLS and the simulation is of the order of 1\,$\sigma$ of the field-to-field variance (blue shaded region on Fig.\,\ref{fig:num_simu_in_vs_out}). Field-to-field variance could thus explain the small offset between the N2CLS counts in the two fields.

    \begin{table*}
        \centering
        \begin{tabular}{cccclcccc}\toprule
            \multicolumn{4}{c}{\textbf{1.2\,mm}}& &\multicolumn{4}{c}{\textbf{2\,mm}}
            \\\cmidrule(r){1-4}\cmidrule(r){6-9}
            S & dN/dS& S$_\mathrm{min}$ & N($\mathrm{>}$S) &  & S & dN/dS & S$_\mathrm{min}$ & N($\mathrm{>}$S) \\
            mJy & deg$^\mathrm{-2}$mJy$^\mathrm{-1}$ & mJy & deg$^\mathrm{-2}$ &  & mJy & deg$^\mathrm{-2}$mJy$^\mathrm{-1}$ & mJy & deg$^\mathrm{-2}$ \\\midrule
            0.66 & 5153$^\mathrm{+2575}_\mathrm{-2227}$ & 0.60 & 2546$^\mathrm{+317}_\mathrm{-378}$ &  & 0.24 & 7294$^\mathrm{+3388}_\mathrm{-2961}$ & 0.20 & 1134$^\mathrm{+207}_\mathrm{-211}$ \\
            0.81 & 3337$^\mathrm{+1377}_\mathrm{-1090}$ & 0.73 & 1879$^\mathrm{+238}_\mathrm{-268}$ &  & 0.33 & 2653$^\mathrm{+1195}_\mathrm{-1039}$ & 0.27 & 593$^\mathrm{+121}_\mathrm{-132}$ \\
            0.98 & 2100$^\mathrm{+821}_\mathrm{-696}$ & 0.88 & 1358$^\mathrm{+187}_\mathrm{-211}$ &  & 0.45 & 1237$^\mathrm{+581}_\mathrm{-449}$ & 0.38 & 320$^\mathrm{+83}_\mathrm{-89}$ \\
            1.19 & 1157$^\mathrm{+494}_\mathrm{-417}$ & 1.07 & 959$^\mathrm{+152}_\mathrm{-189}$ &  & 0.61 & 556$^\mathrm{+171}_\mathrm{-137}$ & 0.52 & 171$^\mathrm{+31}_\mathrm{-30}$ \\
            1.44 & 702$^\mathrm{+319}_\mathrm{-287}$ & 1.30 & 695$^\mathrm{+130}_\mathrm{-161}$ &  & 0.84 & 150$^\mathrm{+58}_\mathrm{-48}$ & 0.71 & 67.3$^\mathrm{+16.5}_\mathrm{-15.9}$ \\
            1.75 & 554$^\mathrm{+154}_\mathrm{-123}$ & 1.58 & 624$^\mathrm{+64}_\mathrm{-63}$ &  & 1.15 & 42.1$^\mathrm{+22.8}_\mathrm{-19.8}$ & 0.97 & 28.0$^\mathrm{+10.9}_\mathrm{-9.3}$ \\
            2.12 & 405$^\mathrm{+95}_\mathrm{-85}$ & 1.91 & 418$^\mathrm{+45}_\mathrm{-46}$ &  & 1.58 & 18.4$^\mathrm{+16.5}_\mathrm{-12.5}$ & 1.33 & 14.2$^\mathrm{+10.1}_\mathrm{-7.2}$ \\
            2.57 & 224$^\mathrm{+56}_\mathrm{-48}$ & 2.32 & 252$^\mathrm{+33}_\mathrm{-33}$ &  & 2.16 & 5.24$^\mathrm{+9.23}_\mathrm{-4.16}$ & 1.82 & 5.15$^\mathrm{+8.86}_\mathrm{-4.39}$ \\
            3.12 & 111$^\mathrm{+32}_\mathrm{-26}$ & 2.82 & 142$^\mathrm{+24}_\mathrm{-22}$ &  &  &  &  &  \\
            3.79 & 52.2$^\mathrm{+19.8}_\mathrm{-17.3}$ & 3.42 & 73.2$^\mathrm{+17.8}_\mathrm{-17.7}$ &  &  &  &  &  \\
            4.59 & 16.1$^\mathrm{+12.2}_\mathrm{-9.9}$ & 4.15 & 35.2$^\mathrm{+13.4}_\mathrm{-11.5}$ &  &  &  &  &  \\
            5.58 & 6.02$^\mathrm{+7.48}_\mathrm{-5.07}$ & 5.04 & 21.3$^\mathrm{+11.5}_\mathrm{-9.1}$ &  &  &  &  &  \\
            6.76 & 6.48$^\mathrm{+5.93}_\mathrm{-4.26}$ & 6.11 & 15.5$^\mathrm{+10.9}_\mathrm{-7.8}$ &  &  &  &  &  \\
            8.21 & 3.98$^\mathrm{+6.74}_\mathrm{-3.16}$ & 7.42 & 6.46$^\mathrm{+10.73}_\mathrm{-5.60}$ &  &  &  &  &  \\

            \bottomrule
        \end{tabular}
        \caption{Combined differential and cumulative galaxy number counts at 1.2 and 2\,mm from the observation on the two fields. The column notations are similar as in Table\,\ref{tab:1.2_source}.}\label{tab:combined_gal}
    \end{table*} 

   \begin{figure*}[tb]
   \centering
      \includegraphics[clip=true,width=1.0\textwidth]{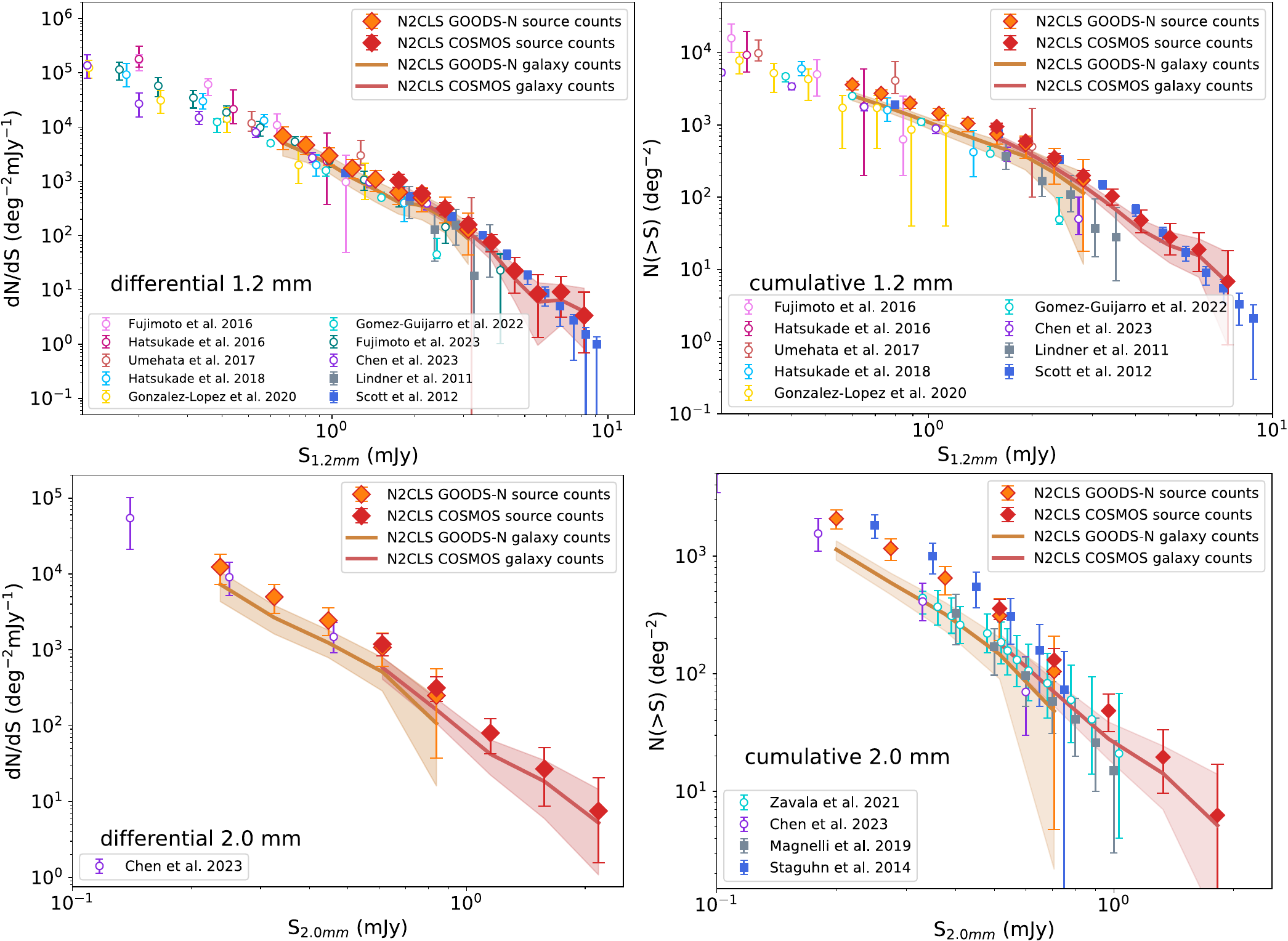}
      \caption{Comparison between N2CLS GOODS-N (orange diamonds) and COSMOS (red diamonds) source number counts  at 1.2\,mm (top panels) and 2\,mm (bottom panels). Differential and cumulative number counts are presented in the left and right panels, respectively. In each panel, the N2CLS galaxy number counts (see Sect.\,\ref{sc:numbercounts_source_to_gal_corr}) and the corresponding 1\,$\sigma$ confidence interval are represented using red (COSMOS) and orange (GOODS-N) solid lines and shaded regions. All results from interferometric observations at 1.1 or 1.2\,mm \citep{Fujimoto+16,Hatsukade+16,Umehata+17,Hatsukade+18,Gonzalez-Lopez+20,Gomez-Guijarro+22,Chen+23,Fujimoto+23} and 2\,mm \citep{Zavala+21,Chen+23} are shown as open circles. The measurements from single-dish observations at 1.1 or 1.2\,mm \citep{Lindner+11,Scott+12} and 2\,mm \citep{Staguhn+14,Magnelli+19} are represented by filled squares.}
      \label{fig:numcounts_vs_obs}
   \end{figure*}

\subsection{From source to galaxy number counts}\label{sc:numbercounts_source_to_gal_corr}

In Sect.\,\ref{sc:numbercounts_calc_validation}, we showed that we are able to measure reliably the source counts from the N2CLS data. However, as shown in Fig.\,\ref{fig:num_simu_in_vs_out}, the galaxy number counts in the simulation (black dashed line) are lower than the source number counts (blue solid line). This difference is mainly due to the blending of several sources inside the $\sim$10-30" beam of single-dish instruments \citep{Hayward+13,Cowley+15,Scudder+16}. This effect has been extensively studied using the SIDES simulation in \citet{Bethermin+17}.\\

We use the SIDES simulation to compute the conversion factor to apply to the source counts to derive the galaxy counts. This multiplicative correction factor is called R$_{\rm diff}$ for the differential number counts and R$_{\rm cum}$ for the cumulative number counts. It is estimated using the ratio between the SIDES intrinsic galaxy number counts and the SIDES source counts derived from the noiseless blob catalog. The values of these corrections are summarize in Table\,\ref{tab:1.2_source} and Table\,\ref{tab:2.0_source}.\\

Finally, we derive the mean of the galaxy number counts in the flux density range where data from the two fields overlap. We use an inverse-variance weighted average of galaxy number counts in each field. For the bright and faint end, we directly use the measurements in COSMOS and GOODS-N field, respectively. The values and uncertainties of the combined 1.2 and 2\,mm galaxy number counts are given in Table~\ref{tab:combined_gal}.\\

\subsection{N2CLS number counts and comparison with previous observations}\label{sc:numbercounts_results}

\subsubsection{Internal consistency of N2CLS number count }

As demonstrated in the previous sections, we can reliably derive source number counts from our data. Before comparing with other measurements in the literature, we perform a last consistency check by comparing the number counts from our two fields in the flux density regimes where they overlap. In Fig.\,\ref{fig:numcounts_vs_obs}, we show the differential and cumulative  number counts from N2CLS at 1.2 and 2\,mm, together with a large compilation from the literature. The two N2CLS fields agree at the 1\,$\sigma$ level at both wavelengths.

\subsubsection{Comparison with other 1.1 and 1.2\,mm number count measurements}

\label{sc:1mm_comp}

The source number counts around 1\,mm have been widely explored in the literature. Before ALMA, observational constraints were obtained from single-dish surveys with either AzTEC/JCMT+ASTE (see \citealt{Scott+12} for a compilation of all the fields) or MAMBO/30m \citep{Lindner+11}. When ALMA reached its full capacity, new deeper but narrower surveys started, providing constraints on the sub-mJy regime \citep{Oteo+16, Aravena+16,Fujimoto+16,Hatsukade+16,Umehata+17,Franco+18,Hatsukade+18,Gonzalez-Lopez+20,Gomez-Guijarro+22,Chen+23,Fujimoto+23}. We show in Fig.\,\ref{fig:numcounts_vs_obs} a comparison of our results with previously published works. \\

We apply a corrective factor to 1.1\,mm data to allow a direct comparison with our 1.2\,mm survey. We use a value of 0.8 for 1.2mm-to-1.1mm flux ratio computed from the main-sequence SED template from \citet{Bethermin+17} at z=2, which is both the median redshift of $\sim$1\,mJy sources expected from models \citep[e.g.,][]{Bethermin+15b} and measured for slightly fainter \citep{Gomez-Guijarro+22} or brighter samples \citep{Brisbin+17}. For both the differential and cumulative number counts, we multiply the x-axis flux by this factor of 0.8. Contrary to the cumulative number counts, the differential number counts (y-axis) are flux dependent and we thus divide them by the 0.8 to take this into account.\\

Our measurements agree with the previous single-dish surveys within 1\,$\sigma$. Our differential source number counts measurements reach deeper flux densities by a factor of 2 and 4 than the previous generation of single-dish surveys by \citet{Scott+12} and \citet{Lindner+11}, respectively. We explore for the first time the sub-mJy regime in a blank field with a single dish, bridging the gap between sub-mJy interferometric constraints and the $>$1\,mJy single-dish surveys.\\

Our source number counts measurements are marginally high compared to the bulk of the interferometric number counts from ALMA. However, after applying the source-to-galaxy corrective factor to our number counts to obtain the galaxy number counts (R$_\mathrm{diff}$ and R$_\mathrm{cum}$, see Sect.\,\ref{sc:numbercounts_source_to_gal_corr}), both ALMA and N2CLS results agrees very well. This highlights that the resolution effects must be taken into account to interpret millimeter deep surveys.

GOODS-N is known to contain several DSFGs associated with galaxy overdensities, i.e. HDF 850.1 at z=5.183 \citep{Walter+12,ArrabalHaro+18} and GN20 at z=4.05 \citep{Daddi+09}. However, we do not observe any significant excess of 1.2\,mm galaxy number counts compared to ALMA measurements in other fields. This could be due to the dilution by the dominant population of field DSFGs that have a much wider range of redshifts. Besides, recent studies reveal that other members of these overdensities have orders of magnitude lower SFR and IR luminosity than the known DSFGs \citep[e.g][]{Calvi+21}, making them unlikely to be detected by the N2CLS survey. Thus, we do not expect our millimeter number counts from the smaller GOODS-N field to be significantly biased by the known overdensities. 

\subsubsection{Comparison with other 2\,mm number count measurements}

In the past decades, only a few blind surveys at around 2\,mm have been carried out. Two surveys have been performed using the Goddard IRAM Superconducting Millimeter Observer (GISMO) camera at the focus of the IRAM 30-meter telescope in the GOODS-N \citep[31\,arcmin$^2$,][]{Staguhn+14} and COSMOS \citep[250\,arcmin$^2$,][]{Magnelli+19}. Since they are taken with the same telescope, these single-dish data have thus a similar beam size, and the source counts can be compared directly. Two ALMA surveys also determined the number counts: the MORA 2\,mm survey that mostly overlaps with the CANDELS stripe in COSMOS (184\,arcmin$^2$, \citealt{Zavala+21}), and the ALMACAL archival survey based on ALMA calibrator observations (157\,arcmin$^2$, \citealt{Chen+23}). These two interferometric surveys have a sub-arcsec resolution, and we can assume that they measure directly the galaxy number counts.\\

In the bottom panels of Fig.\,\ref{fig:numcounts_vs_obs}, we show a comparison between the number counts from these surveys and N2CLS. Our new N2CLS measurements agree with previous GISMO GOODS-N measurements from \citet{Staguhn+14}. The N2CLS probes slightly fainter fluxes, and have similar error bars for GOODS-N at faint fluxes, which may seem surprising considering our $\sim$5 larger survey area. This could be explained by our propagation of the flux deboosting uncertainties to the final error bars. There is a mild systematic offset ($<$2$\sigma$) between the COSMOS number counts from GISMO \citep{Magnelli+19} and NIKA2. Our survey covers a 4.4 times larger area, and we cannot exclude that the area used by the GISMO study is not representative of the full field as suggested by the absence of sources in the eastern part of their map. Unfortunately, our fields only partially overlap preventing a N2CLS measurements of the counts in the same exact region.\\

As explained in Sect.\,\ref{sc:numbercounts_source_to_gal_corr}, the galaxy number counts measured by interferometers cannot be directly compared with source counts from single-dish surveys. Before comparing ALMA observations with N2CLS, we applied a corrective factor to our number counts (R$_\mathrm{diff}$ and R$_\mathrm{cum}$, see Sect.\,\ref{sc:numbercounts_source_to_gal_corr}). The galaxy number counts obtained after these corrections (orange and red solid lines in Fig.\,\ref{fig:numcounts_vs_obs}) agrees very well with the ALMA data, highlighting the importance to take into account resolution effects in the (sub-)millimeter. We can also note that N2CLS is deeper and covers a larger area than the MORA survey, demonstrating the efficiency of single-dish telescopes to perform wide and deep millimeter surveys.\\

Overall, our measurements agree with the literature, except a mild tension with GISMO \citep{Magnelli+19} measurements. Thanks to the mapping speed of the NIKA2 camera, our survey covers the full range explored by the previous surveys in a homogeneous way.

   \begin{figure*}[tb]
   \centering
      \includegraphics[clip=true,width=.99\textwidth]{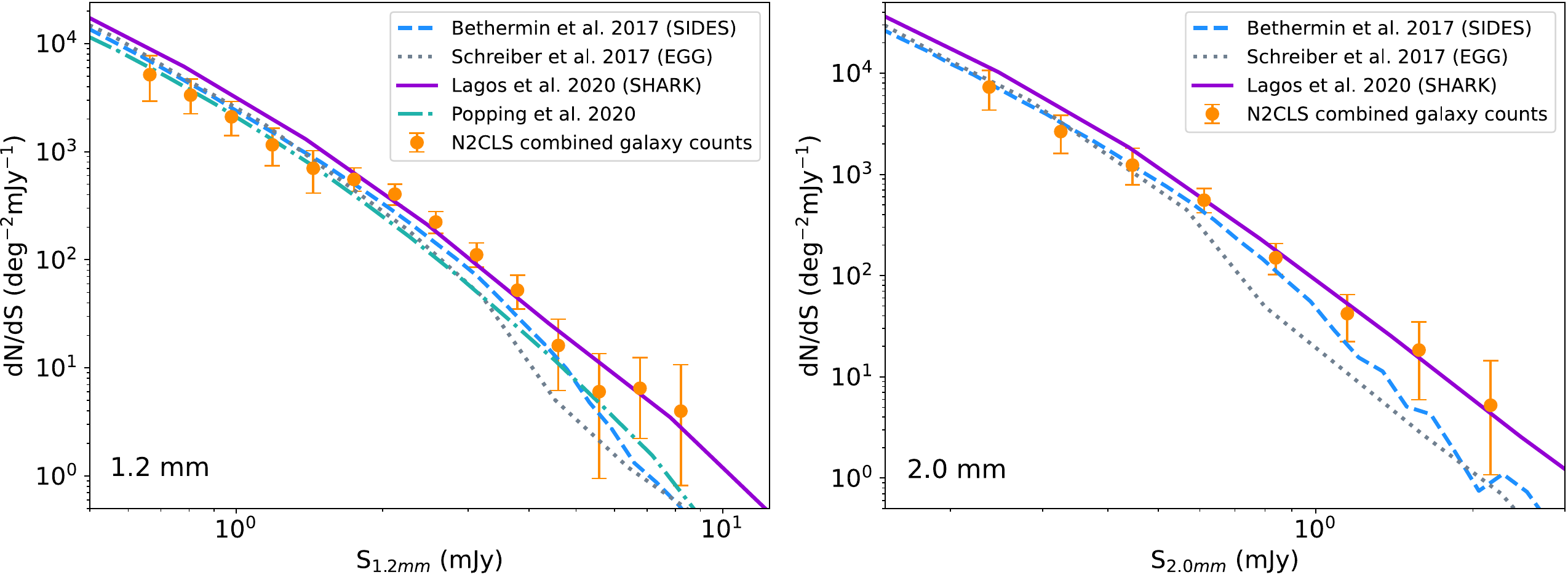}
      \caption{Comparison between the N2CLS differential galaxy number counts (both fields combined, see Sect.\,\ref{sc:numbercounts_source_to_gal_corr}) and the predictions from semi-empirical \citep{Bethermin+17,Schreiber+17,Popping+20} and semi-analytical \citep{Lagos+20} models.}
      \label{fig:galcounts_vs_model}
   \end{figure*}

\subsection{Comparison with models}\label{sc:numbercounts_modelcomp}

We also compared our new measurements with number counts predictions from various models of galaxy evolution, including both semi-empirical \citep{Bethermin+17,Schreiber+17,Popping+20} and semi-analytical \citep{Lagos+20} models.\\

The Simulated Infrared Dusty Extragalactic Sky (SIDES, \citealt{Bethermin+17}) and the Empirical Galaxy Generator (EGG, \citealt{Schreiber+17}) start from the stellar mass function and the evolution of the star-forming main sequence from observations to predict infrared and (sub-)millimeter fluxes. They also separately account for the emission of main-sequence and starbursts galaxies using different SED templates, both evolving with redshift.
The semi-empirical model of \citet{Popping+20} assigns star formation rates in dark matter halos following the SFR-halo relation from the \textsc{UNIVERSEMACHINE} \citep{Behroozi+19}, and then uses empirical relations to estimate the dust mass and obscured star formation to predict the millimeter fluxes. Their prediction are converted from 1.1 to 1.2\,mm using the method explained in Sect.\,\ref{sc:1mm_comp}.\\

The \textsc{SHARK} semi-analytical model is introduced in \citet{Lagos+18}. This type of models applies semi-analytical recipes to model the evolution of galaxies in dark-matter halos from numerical simulations. The dust emission and (sub-)millimeter fluxes of galaxies is then predicted based on their gas content and metallicity using a framework described in \citet{Lagos+19,Lagos+20}. Since, their original number counts are cumulative, we differentiated their curves to obtain the differential ones. \\

In Fig.~\ref{fig:galcounts_vs_model}, we show the comparison between model predictions and our N2CLS results. Since most models are predicting galaxy rather than source number counts, we use the combined galaxy number counts from our two fields derived in Sect.\,\ref{sc:numbercounts_source_to_gal_corr}. At 1.2\,mm, the three semi-empirical models (SIDES, EGG, and \citealt{Popping+20}) agree at the 1\,$\sigma$ level with our observations at the faint end ($<$1.5\,mJy), but tend to be systematically lower at the bright end ($>$1.5\,mJy, $\sim$1.5\,$\sigma$ for SIDES, up to 3\,$\sigma$ for EGG and \citealt{Popping+20}). In contrast, the \textsc{SHARK} semi-analytical model is compatible at the bright end, but has a systematic 1.5\,$\sigma$ excess at the faint end. A similar behavior is observed at 2\,mm. However, because of the larger uncertainties on the measurements, SIDES and \textsc{SHARK} remain compatible with our observations at $\sim$1\,$\sigma$. The EGG model remains significantly under-predicted at the bright end ($>$0.6\,mJy). This difference of behavior between semi-empirical and semi-analytical models is discussed in Sect.\,\ref{sc:discussion}.\\

\section{Discussion}\label{sc:discussion}

\subsection{Modeling the millimeter number counts}

{As shown in Sect.\,\ref{sc:numbercounts_modelcomp}, recent models are all able to reproduce the main trend of the millimeter number counts. The tension between models and observations remains small. This suggests that minor adjustments may be sufficient to reach a full agreement. Considering how challenging the (sub-)millimeter number counts have been for semi-analytical models and hydrodynamical simulations during the last two decades \citep[e.g.][]{Baugh+05,Hayward+13,Cousin+15,Somerville+15,Narayanan+15,Lacey+16,Hayward+21}, this highlights the impressive progress made during the recent years. The small residual tension between \textsc{SHARK}  \citep{Lagos+20} and observations at the faint end ($<$1.5\,mJy) could be solved by a small tuning of the star formation or feedback recipes. However, considering the large number of degrees of freedom in this type of models, it is hard to predict which exact change is the most relevant.\\

Semi-empirical models are more flexible, and updated models were often proposed shortly after the delivery of new observational constraints \citep[e.g.,][]{Chary+01,Franceschini+01,Lagache+04,Bethermin+11,Lapi+11,Gruppioni+13,Casey+18}. These updates showed that modifications of the evolution of dusty galaxy populations or of their SEDs were necessary. The three recent semi-empirical models discussed in Sect.\,\ref{sc:numbercounts_modelcomp} are all slightly lower than our measured number counts at the bright end ($>$1.5\,mJy). This systematic trend could be clue of a common problem. The EGG \citep{Schreiber+17} and \citet{Popping+20} models do not include the effect of strong lensing on the number counts \citep[e.g.][]{Negrello+10}. This explains why these two models are lower than the SIDES model \citep{Bethermin+17} in this regime, since 3\,\%, 10\,\%, and 60\,\% of sources above 1.5\,mJy, 3\,mJy, and 5\,mJy, respectively, are strongly lensed in SIDES.\\

However, even taking into account the lensing, SIDES number counts remain marginally low at the bright end ($>$1.5\,mJy). The two most simple explanations are a lack of galaxies with cold dust SEDs leading to fainter millimeter fluxes, and a small deficit of galaxies with high SFR (SFR$\gtrsim$500\,M$_\odot$/yr). For instance, the fraction of starbursts in SIDES is fixed to 3\,\% at z$>$1 whatever the stellar mass. A slightly higher fraction of starbursts at high mass could be sufficient to match the observations. But hydrodynamical simulations find that high-z major mergers may be less efficient to enhance star formation \citep{Fensch+17}. In addition, the SIDES model has a sharp SFR limit of 1000\,M$_\odot$/yr. This limit was motivated by the ALMA follow up of bright millimeter sources, which showed that they were breaking into several components \citep{Karim+13}. However, a smoother cut allowing rare SFR$>$1000\,M$_\odot$/yr objects could reduce the tension with the observations.\\

Finally, the millimeter number counts are very sensitive to the assumptions on the far-infrared SEDs. The SED templates used by SIDES \citep{Bethermin+15} and EGG \citep{Schreiber+18} have been calibrated using the observed mean evolution of the dust temperature up to z$>$4, and are compatible with most of the recent ALMA results \citep{Bethermin+20,Faisst+20,Sommovigo+22}. A recent stacking-based analysis from \citet{Viero+22} suggests even higher dust temperatures, which would lead to an even larger disagreement between observed counts and empirical models. In contrast, recent studies also discovered DSFGs with far-infrared SEDs peaking at a significantly longer wavelength than those of normal star-forming galaxies (e.g. \citealt{Jin+19}). At a fixed star formation rate, sources with these apparently cold dust SEDs have higher millimeter fluxes. In any case, a larger scatter in the dust properties would naturally lead to higher number counts at the bright end. Extensive follow-up campaigns of millimeter sources with well-controlled selection biases might be the key to properly calibrate the SEDs in the models.

\subsection{A framework for accurate interpretation of single-dish millimeter data}

We highlighted the difference between the number density of sources viewed by high angular resolution ($\sim$1"), interferometric observations and low resolution ($\sim$15"), single-dish observations in the millimeter. The impact of angular resolution on flux measurements in single-dish observations has been previously discussed in interferometric follow up \citep[e.g.][]{Karim+13,Simpson+20} and modeling papers \citep[e.g.][]{Hayward+13,Scudder+16,Bethermin+17}. Our work, for the first time, quantitatively estimates and corrects for this effect for a single-dish blind survey. As shown in Sect.\,\ref{sc:numbercounts_source_to_gal_corr}, the differences between the galaxy number counts and source number counts can reach a factor of two at 2\,mm, even with a 30\,m telescope. Correcting for this effect (Sect.\,\ref{sc:numbercounts_results}), we showed that interferometric and single-dish observations are fully consistent. Our paper proposes a new framework to interpret single-dish number counts without requiring for systematic follow-up observations, and which can be applied to future surveys \citep[e.g.,][]{Wilson+20, Klaassen+20,Ramasawmy2022}.

\subsection{Modeling the number of sources detected only at 2\,mm}

\label{sect:2mmonly_model}

In Sect.\,\ref{sc:purity}, we found that a large fraction of the N2CLS sources are detected only at 2\,mm (when considering S/N thresholds corresponding to 80\% purity). This could suggest a large population of galaxies at very high redshifts. In contrast, in the SIDES input catalog, we found only an average of less than one source per field in GOODS-N and COSMOS, respectively, which are below the survey flux limit at 1.2\,mm and above it at 2\,mm. This corresponds to S$_{\rm 1.2\,mm}$$<$0.4\,mJy and S$_{\rm 2.0\,mm}$$>$0.1\,mJy in GOODS-N, and S$_{\rm 1.2\,mm}$$<$1.7\,mJy and S$_{\rm 2.0\,mm}$$>$0.5\,mJy in COSMOS. The observations seem to disagree strongly with the SIDES model. However, the instrument noise can be responsible for these sources. For instance, a source intrinsically just above the detection limit in both band will be detected only at 2\,mm if it is on a negative fluctuation of the noise at 1.2\,mm. A similar phenomenon has already been identified for red \textit{Herschel}/SPIRE sources \citep{Bethermin+17,Donevski+18}.\\

We used the 117 end-to-end simulations of each field presented in Sect.\,\ref{sc:numbercounts_calc_validation} to investigate the nature of these 2\,mm-only sources. We found an average of 34 and 25 sources per field in GOODS-N and COSMOS, respectively (high-quality region only, see red contours in Fig.\,\ref{fig:n2cls_1.2mm_maps} and \ref{fig:n2cls_2.0mm_maps}). Assuming Poisson uncertainties, this is compatible at 2\,$\sigma$ with the 23 and 27\,sources detected in the real data (Sect.\,\ref{sc:purity}). In both fields, 87\,\% of these sources are associated to a counterpart in the blob catalog. Most of these 2\,mm detections are thus associated to objects present in the simulation, and are not pure noise artifacts. This suggests that the combination of instrument noise, data reduction pipeline and source extraction procedures could produce this apparent excess of 2\,mm-only sources.\\

We checked if increasing the S/N threshold improves the situation. For a S/N threshold corresponding to 95\,\% purity, we have 2 and 11 sources in the real GOODS-N and COSMOS catalogs, while the end-to-end simulations contain in average 6 and 8 sources per field. More than 98\,\% of these sources are associated to an object in the blob catalog. However, there is still a mismatch with the input catalog in which less than one source per field is detected. This is not surprising, since we increased both 1.2 and 2\,mm S/N thresholds and the mechanism producing spurious 2\,mm-only detections still apply.\\

As shown by our simulations, the selection of sources detected only a 2\,mm by NIKA2 is thus not a reliable way to select very high-redshift candidates. This also highlights the importance of end-to-end simulation to properly compare models with observations.

\section{Conclusion}\label{sc:conclusion}
We presented the first results of the NIKA2 Cosmological Legacy Survey (N2CLS), a large blind millimeter survey in the GOODS-N and COSMOS fields with the NIKA2 camera on the IRAM 30\,m telescope. We used the NIKA2 observations from October 2017 to May 2021, representing 86.15\,h and 84.7\,h on field for GOODS-N and COSMOS, respectively. The area used in our analysis is 159\,arcmin$^2$ for GOODS-N and 1010\,arcmin$^2$ for COSMOS. The survey reaches an unprecedented combination of depth and sky coverage at 1.2 and 2\,mm. The main steps of our analysis and our main results are summarized below.

\begin{itemize}

    \item We built the maps using the IRAM PIIC software \citep{Zylka13}, and extracted the sources using our custom \textit{nikamap} \citep{beelen_alexandre_2023_7520530} package based on \textit{Astropy} \citep{astropy:2013, astropy:2018, astropy:2022} and  \textit{Photutils} \citep{larry_bradley_2022_6825092}. \\

	\item To characterize the performance of our analysis pipeline, we performed 117 end-to-end simulations of each field based on the SIDES model of galaxy evolution \citep{Bethermin+17,Gkogkou+22}. They take advantage of the simulation mode of the PIIC pipeline, which accepts SIDES maps as input models to be injected in real NIKA2 timeline data. A half-difference method was applied to these timelines to remove only the true astrophysical signal but not the injected one. Maps and catalogs from the simulations were then produced identically to N2CLS data.\\

    \item We then compared the output source catalogs of these end-to-end simulations with the input ones to determine the performance of our source extraction algorithm. Because of the angular resolution of NIKA2, we use for the input catalogs the blobs extracted from the noiseless maps rather than the individual galaxies. For each field and wavelength, we determined the sample purity as a function of the S/N threshold, and the completeness as a function of the source flux and the local noise level.  With the S/N thresholds of 80\% purity, we detect 120 and 195 sources at 1.2\,mm in GOODS-N and COSMOS respectively, and 67 and 76 sources at 2\,mm. In the 1.2\,mm maps of GOODS-N and COSMOS, we detect sources as faint as 0.4\,mJy and 1.7\,mJy in uncorrected PSF fluxes. At 2\,mm, we reach limiting uncorrected PSF fluxes of 0.1\,mJy and 0.5\,mJy in GOODS-N and COSMOS, respectively. \\
 
     \item We also computed the ratio between the output (measured) and the input (simulated) flux densities, taking into account the effects of both data reduction (flux filtering) and source extraction (flux boosting). The measured flux densities are on average lower than the input ones in GOODS-N, demonstrating that some flux is lost during the map making and providing us the corrections to apply.\\

    \item We then computed the source number counts after correcting for all the effects listed above. We checked using our end-to-end simulations that our method is accurate. In addition, we derived the correction to convert our source number counts into galaxy number counts. This correction is necessary to compare our results with ALMA measurements and with models.\\

    \item At 1.2\,mm, our measurements cover the full flux density range from previous single-dish surveys and goes a factor of 2 deeper, reaching the sub-mJy regime. We probe in an homogeneous way 1.5 orders of magnitude in flux density, and connect the bright single-dish number counts to the deep interferometric number counts. Our new measurements agree well with previous measurements after taking into account the resolution effects.\\

    \item At 2\,mm, our measurements match the depth of the deepest interferometric number counts and extend a factor of 2 above the brightest constraints. Our results agree with the single-dish measurements from \citet{Staguhn+14}, and also with the interferometric constraints from \citet{Zavala+14} and \citet{Chen+23} after correcting for resolution effects. Results from \citet{Magnelli+19} are systematically $\sim$1-$\sigma$ lower than our measurements\\

    \item Finally, we compared our measured galaxy number counts with a selection of recent semi-empirical \citep{Bethermin+17,Schreiber+17, Popping+20} and semi-analytical \citep{Lagos+20} models. The semi-empirical models agree at low flux density ($<$1.5\,mJy), but tend to under-predict the counts at bright flux density ($>$1.5\,mJy). We discussed several possible causes such as the lack of strong lensing in some models, a deficit of high-SFR galaxies, or too few objects with a cold dust SEDs. In contrast, the semi-analytical model of \citet{Lagos+20} over-predicts the counts at low flux, while agreeing at higher flux.

\end{itemize}

    The measurements and the models of millimeter source counts are now close to converge. Stronger constraints will come from a full characterization of these sources, and will allow us to test our models in greater details. The upcoming follow-up observations with NOEMA/ALMA will pinpoint the location of galaxies contributing to the observed N2CLS flux. The rich ancillary data and ongoing JWST observations like COSMOS-Web \citep{Casey+22} will help identify the multi-wavelength counterparts of N2CLS sources, construct their full SED, and determine their redshift distribution and physical properties. Thanks to its volume-complete flux selection, the N2CLS sample is an ideal reference sample to perform this full characterization of DSFGs.\\

    We expect to reach a depth 1.5 deeper in COSMOS compared to this work. The final COSMOS catalogs will be released together with the full COSMOS data release in a forthcoming paper. The identification and properties sources with SNR$>4$ in GOODS-N will be detailed in Berta et al. (in prep) together with the complete source catalog. Also in GOODS-N, Ponthieu et al. (in prep) will discuss the confusion noise due to distant galaxies for the IRAM 30m telescope.

\begin{acknowledgements}

The authors would like to thank the anonymous referee for the comments on the manuscript, as well as Gerg{\"o} Popping and Claudia Lagos for providing their model prediction on millimeter number counts. 
LB warmly acknowledges financial support from IRAM for his first year of PhD thesis and the support from the China Scholarship Council grant (CSC No. 201906190213).

We acknowledge financial support from the ”Programme National de Cosmologie and Galaxies” (PNCG) funded by CNRS/INSU-IN2P3-INP, CEA and CNES, France, from the European Research Council (ERC) under the European Union's Horizon 2020 research and innovation programme (project CONCERTO, grant agreement No 788212) and from the Excellence Initiative of Aix-Marseille University-A*Midex, a French "Investissements d'Avenir" programme. 

This work is based on observations carried out under project numbers 192-16 with the IRAM 30m telescope. IRAM is supported by INSU/CNRS (France), MPG (Germany) and IGN (Spain).

We would like to thank the IRAM staff for their support during the campaigns. The NIKA2 dilution cryostat has been designed and built at the Institut N\'eel. In particular, we acknowledge the crucial contribution of the Cryogenics Group, and in particular Gregory Garde, Henri Rodenas, Jean-Paul Leggeri, Philippe Camus. This work has been partially funded by the Foundation Nanoscience Grenoble and the LabEx FOCUS ANR-11-LABX-0013. This work is supported by the French National Research Agency under the contracts "MKIDS", "NIKA" and ANR-15-CE31-0017 and in the framework of the "Investissements d’avenir” program (ANR-15-IDEX-02). This work has benefited from the support of the European Research Council Advanced Grant ORISTARS under the European Union's Seventh Framework Programme (Grant Agreement no. 291294). E. A. acknowledges funding from the French Programme d’investissements d’avenir through the Enigmass Labex. A. R. acknowledges financial support from the Italian Ministry of University and Research - Project Proposal CIR01$\_$00010. 

The NIKA2 data were processed using the Pointing and Imaging In Continuum (PIIC) software, developed by Robert Zylka at the Institut de Radioastronomie Millimetrique (IRAM) and distributed by IRAM via the GILDAS pages. PIIC is the extension of the MOPSIC data reduction software to the case of NIKA2 data.

This work made use of Astropy, a community-developed core Python package and an ecosystem of tools and resources for astronomy \citep{astropy:2013, astropy:2018, astropy:2022}. 

This research made use of Photutils, an Astropy package for
detection and photometry of astronomical sources \citep{larry_bradley_2022_6825092}.

\end{acknowledgements}

\bibliographystyle{aa}

\bibliography{biblio}

\begin{appendix}

\section{N2CLS 1.2 and 2\,mm signal maps  \label{sc:N2CLS maps}}


\begin{figure*}[tb]
   \centering
      \includegraphics[clip=true,width=.99\textwidth]{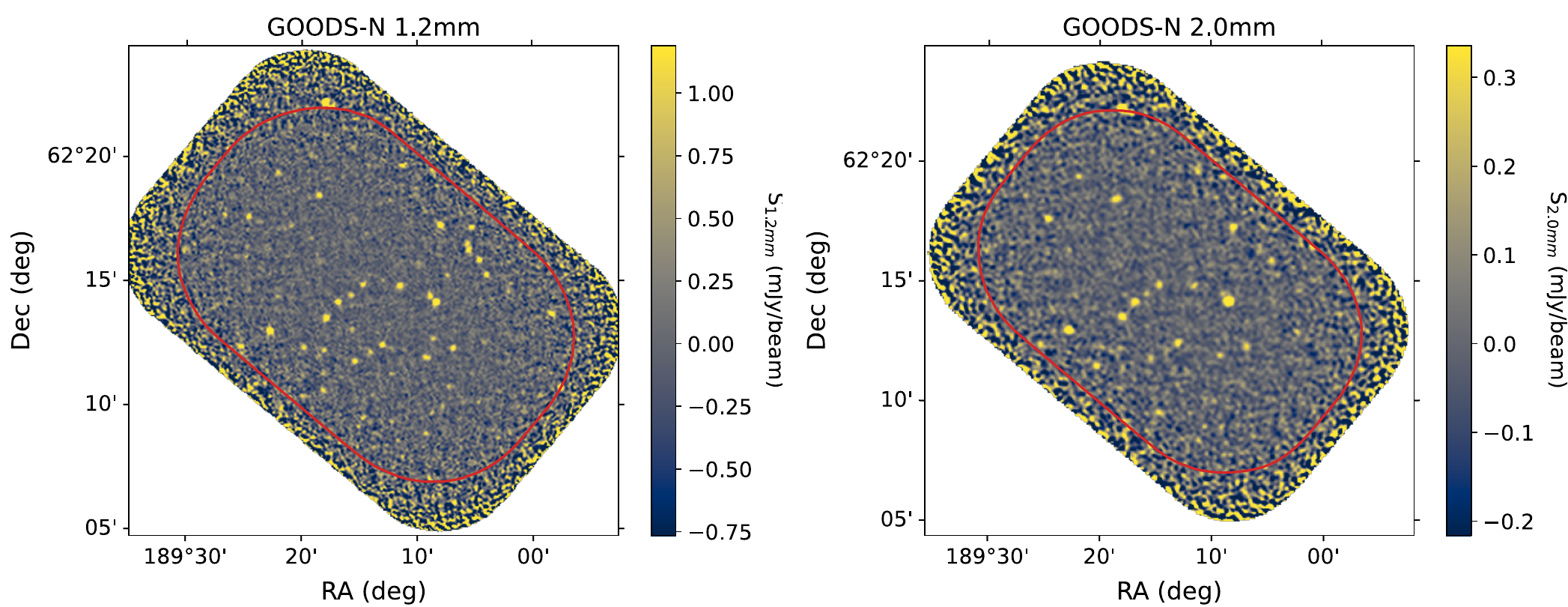}
      \caption{The 1.2\,mm and 2\,mm signal map from N2CLS GOODS-N observations. The high-quality region considered in N2CLS number counts analysis is enclosed by the red contour, which is defined in Sect.~\ref{sc:data_red}.}
      \label{fig:n2cls_gn_signal}
\end{figure*}

\begin{figure*}[tb]
   \centering
      \includegraphics[clip=true,width=.99\textwidth]{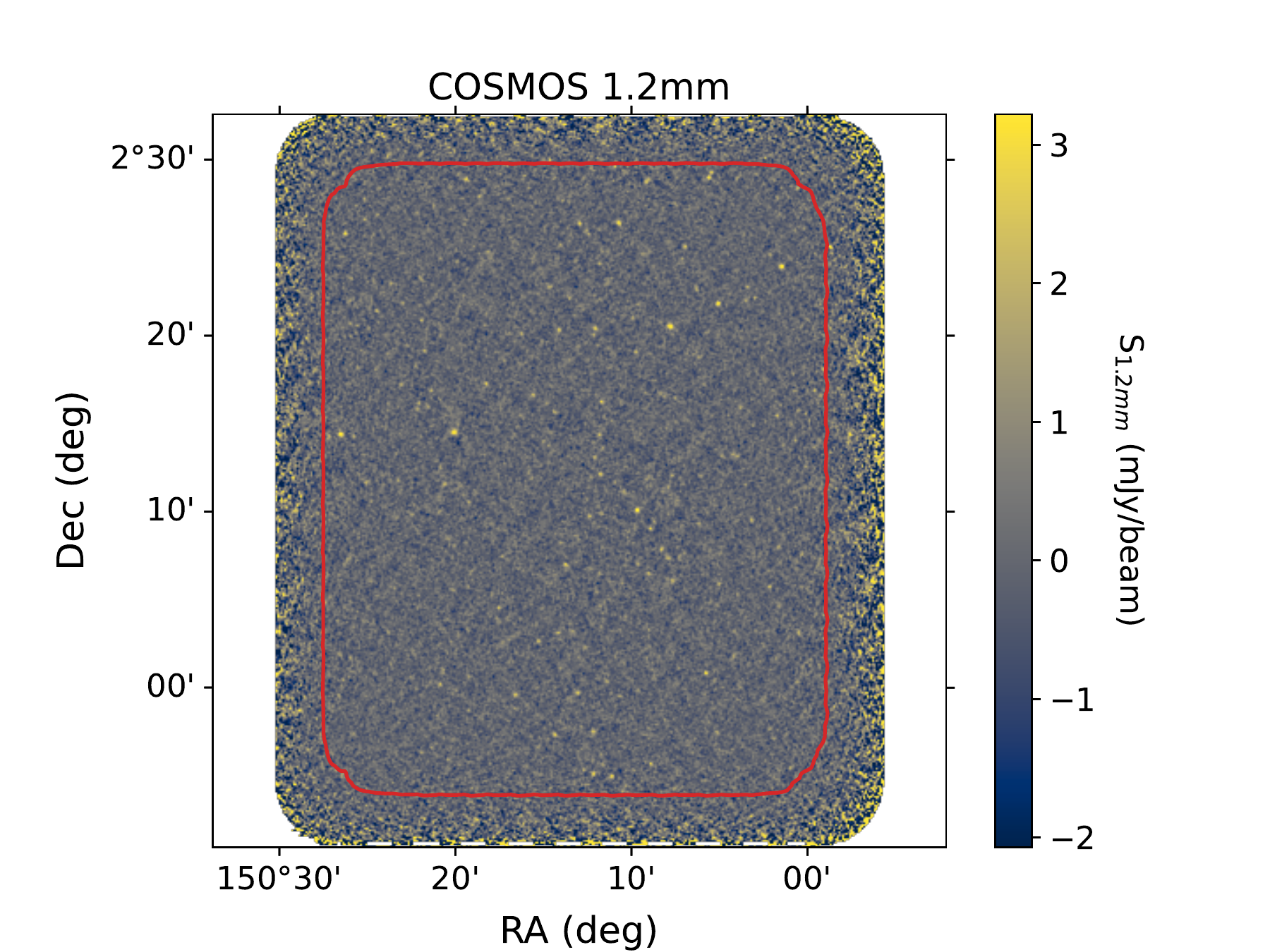}
      \caption{Same as Fig.~\ref{fig:n2cls_gn_signal}. but for N2CLS COSMOS 1.2\,mm observations.}
      \label{fig:n2cls_cos_1.2mm_signal}
\end{figure*}

\begin{figure*}[tb]
   \centering
      \includegraphics[clip=true,width=.99\textwidth]{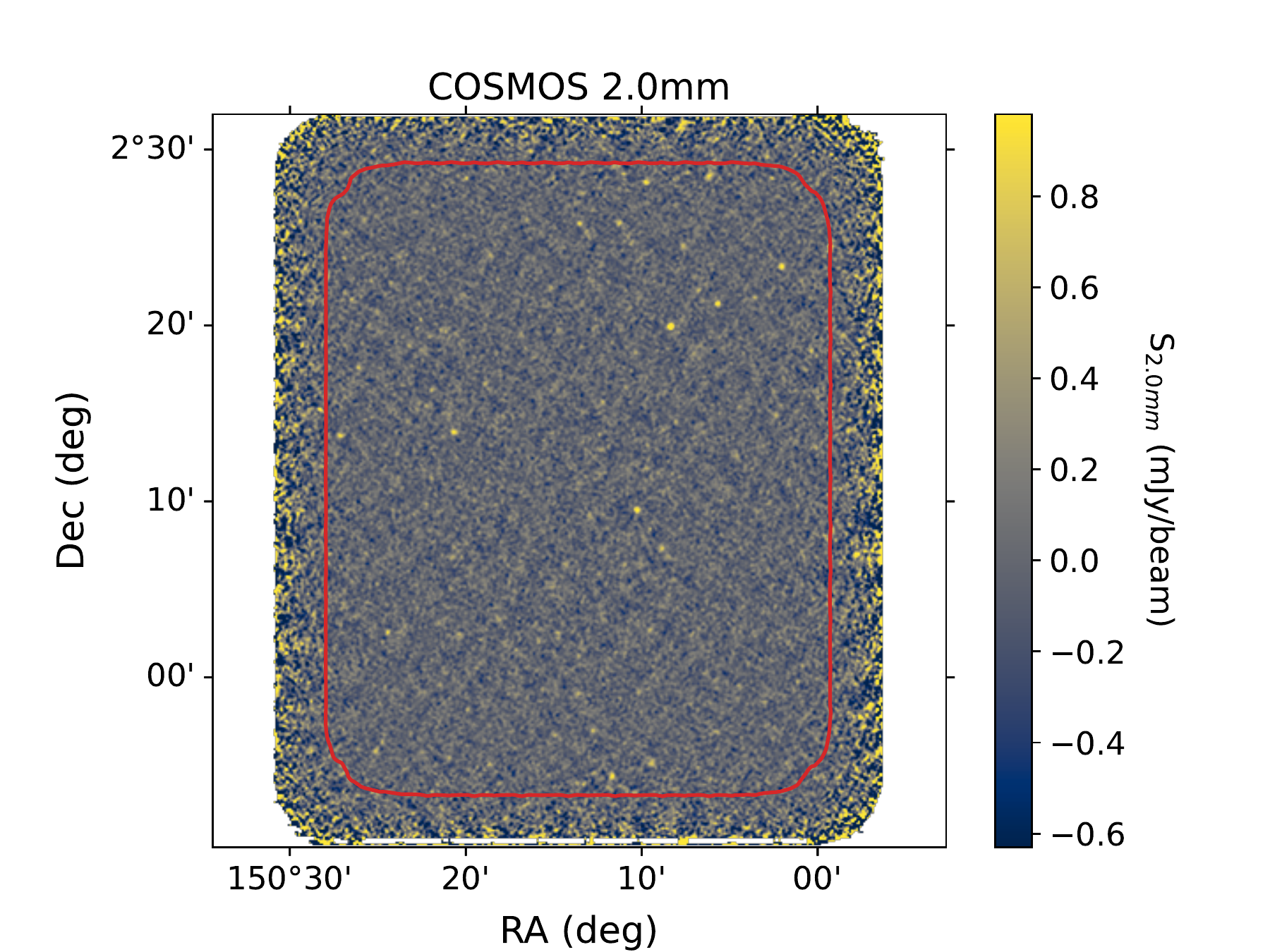}
      \caption{Same as Fig.~\ref{fig:n2cls_gn_signal}. but for N2CLS COSMOS 2\,mm observations.}
      \label{fig:n2cls_cos_2.0mm_signal}
\end{figure*}

\end{appendix}

\end{document}